\documentclass{iopart}
\usepackage{pstricks,pst-node}

\usepackage [dvips]{graphicx}
\usepackage{epsfig}
\usepackage{iopams}

\newcommand{\hf}{\frac12}
\newcommand{\hfs}{{\textstyle{\frac12}}}
\newcommand{\Bigs}{}
\newcommand{\ee}{\mathrm{e}}
\newcommand{\Sbar}{\overline{S}}
\newcommand{\Wint}{W^{(1)}}
\newcommand{\Whf}{W^{(1/2)}}
\newcommand{\Gbar}{\overline{G}}
\newcommand{\Gint}{G^{(1)}}
\newcommand{\Gbarint}{\overline{G}^{(1)}}
\newcommand{\Ghf}{G^{(1/2)}}
\newcommand{\Gbarhf}{\overline{G}^{(1/2)}}
\newcommand{\mub}{\mu_\mathrm{B}}
\newcommand{\mut}{\mu_\mathrm{T}}

\begin{document}

\title{Duality symmetries and effective dynamics in disordered hopping models}
\author{Robert L. Jack}
\address{
Department of Physics, University of Bath,
Bath BA2 7AY, UK}
\author{Peter Sollich}
\address{
King's College London, Department of Mathematics,
London WC2R 2LS, UK}

\begin{abstract}
We identify a duality transformation in one-dimensional hopping models 
that relates propagators in general
disordered potentials linked by an up-down inversion of the energy
landscape. This significantly generalises previous results for
a duality between trap and barrier models.
We use the resulting insights into the symmetries of these
models to develop a real-space renormalisation
scheme that can be implemented computationally
and allows rather accurate prediction of
propagation in these models.  We also discuss the relation
of this renormalisation scheme to earlier analytical
treatments.
\end{abstract}

\maketitle

\section{Introduction}

The motion of particles in disordered environments is important
in many contexts, from 
glass-forming liquids and colloids~\cite{glass_expt,glass_thy}, 
to biomolecules moving in the crowded
environment of the cell~\cite{bio}, to electrical properties 
of disordered materials~\cite{Bern79}.
In this article, we discuss subdiffusive propagation in simple 
one-dimensional models.  While the case of
one-dimensional motion may seem simplistic, it is
relevant for a variety of model systems: from early
studies of electrical transport~\cite{Bern79} to
recently-defined models of glassy behaviour~\cite{BBL}, and also
to the motion of defects
in disordered magnets, to disordered elastic chains
and to networks of resistors and 
capacitors (see \cite{Alex81,HbA87,BouGeo90,MetKla00} for reviews).

The results that we will present 
are based around a duality symmetry which relates
pairs of discrete one-dimensional models.  In a previous 
study~\cite{JS-dual}, 
we showed that motion in apparently disparate models
can be related exactly, at fixed disorder.
Here we generalise those results to a much wider range
of hopping models, by treating their
master equations in a simple operator formalism.
We also discuss how these results are related to
earlier studies of disordered reaction-diffusion
systems by Sch\"utz and Mussawisade~\cite{schutz}.
Based on the symmetries of the problem, we then introduce
a real-space renormalisation scheme in the spirit
of that of le Doussal, Monthus and Fisher~\cite{DMF}: 
our scheme is implemented computationally
and allows rapid prediction of the 
propagation in these energy landscapes for fixed
disorder, at low computational cost.

The central feature of the renormalisation scheme is that on a
given time scale, we have a procedure for decomposing the system into
effective trap and barrier regions.  Particles within trap
regions equilibrate there, while those in barrier regions decay
into the effective traps.  The duality symmetry
relates trap and barrier regions of pairs of models, and demands
that they be treated on an equal footing within the renormalisation
scheme.  In some sense, the renormalisation scheme is connected
with the ideas of an energy landscape in these disordered 
systems~\cite{landscape}, 
but we note that all properties of the energy landscape
are here derived directly from the master operator of the stochastic
dynamics.  In this sense, our scheme allows the
energy landscape to be derived from the dynamical rules of the
system: this is the opposite of the usual situation in which
thermodynamic properties are used to infer the routes by
which dynamical processes take place.  Thus, while our results are
clearly restricted to a very simple class of models, it is natural
to ask if they might be generalised to higher-dimensional energy
landscapes. 

The form of the paper is as follows: in Sec.~\ref{sec:dual} we define
our models and give the duality relation between their
master operators.  The consequences of the duality relation for
propagation in these models are discussed in Sec.~\ref{sec:prop}.
In Sec.~\ref{sec:eff} we explain our effective dynamics scheme;
Sec.~\ref{sec:numerics} contains numerical results for specific ensembles
of disordered models; and Sec.~\ref{sec:outlook} closes with a brief
summary and some open questions.

\section{Models and duality relations}
\label{sec:dual}

We define a disordered one-dimensional hopping model
in terms of rates for hops from site $i$ to sites $i-1$ and $i+1$,
which we denote by $\ell_i$ and $r_i$ 
respectively.  
Let $p_i(t)$ be the probability that a particle
occupies site $i$ at time $t$: the master equation is then
\begin{equation}
\frac{\partial}{\partial t}p_i(t) = \ell_{i+1} p_{i+1}(t)
 + r_{i-1} p_{i-1}(t) - (\ell_i + r_i) p_i(t).
\label{equ:master}
\end{equation}
For concreteness, we consider a periodic chain of $N$ sites,
but we are primarily concerned with propagation of particles
on infinite chains: that is, we consider the limit of large
$N$ before any limit of large time.  In this limit, propagators
will be independent of the choice of boundary conditions.  We also
discuss finite chains with reflecting and absorbing boundaries
in section~\ref{sec:bc} below.

We use an operator
notation where the ket $|i\rangle$ represents the state with
the particle on site $i$, normalised so that
$\langle i|j\rangle=\delta_{ij}$. 
Then, defining the state $|P(t)\rangle=\sum_i p_i(t)|i\rangle$, 
the master equation can be written as
$
\frac{\partial}{\partial t}|P(t)\rangle = \Wint |P(t)\rangle
$
with
\begin{eqnarray}
\Wint
  &=& \sum_{i=1}^{N} 
\big[ \ell_i|i-1\rangle + r_i|i+1\rangle - 
               (l_i+r_i)|i\rangle \big] \langle i| 
\label{equ:Wt1}
\\
  &=& \sum_{i=1}^{N} 
    \big( |i+1\rangle - |i\rangle \big) \big( r_i\langle i| - 
\ell_{i+1}\langle i+1| \big).
\label{equ:Wt}
\end{eqnarray}
In this section, and wherever we consider
systems with periodic boundaries, site $i=N+1$ is equivalent
to site $1$ and site $0$ is equivalent to site $N$. This allows terms
to be rearranged as e.g.\ in going from (\ref{equ:Wt1}) to (\ref{equ:Wt}) above.

We initially focus on an important special case: we assume that
all rates are finite and that
$\prod_i \ell_i = \prod_i r_i$. This ensures that all currents vanish
in the long-time limit of the system.
Under this assumption,
we associate an energy with each site, measured \emph{downwards}
from an arbitrary baseline and 
determined through $\ell_{i+1}\ee^{E_{i+1}}=r_i\ee^{E_i}$. 
Then, the model respects detailed balance with respect to the 
distribution $p_i^\mathrm{eq}=\ee^{E_i}/(\sum_r \ee^{E_r})$,
where the sum runs over all sites.
With this sign convention, a site with large positive
$E_i$ has a large Gibbs weight.  The reason for
this choice will become clear below.

\begin{figure}
\begin{center} \includegraphics[width=6cm]{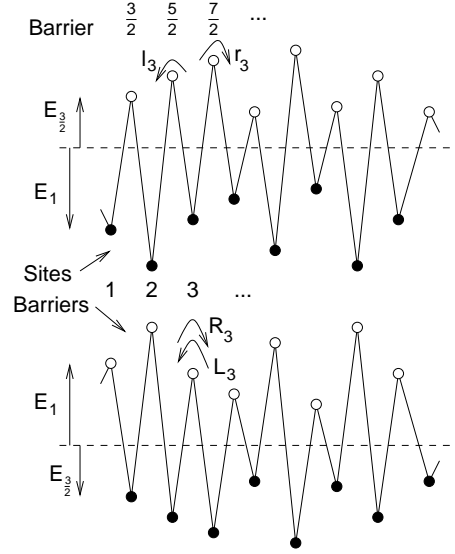} \end{center}
\caption{
(Top) Illustration of an `energy landscape', defined by 
supplementing the site energies $E_i$ by transition
state energies $E_{i+\hf}$.  The rate for
hopping from a site is related to the sum of the
site energy (measured downwards as shown) 
and the adjacent transition state energy (measured upwards).  
(Bottom) On inversion of the potential, the transition
state energies become energies of new sites with
half-integer indices.  Transition rates between
these sites are controlled by the energies $E_i$
which now have an interpretation as transition
state energies.
} \label{fig:invert_all}
\end{figure}

We now discuss general duality relations between pairs of hopping models. 
To this end, we re-parameterise the transition rates $\ell_i$
and $r_i$ by defining transition state energies $E_{i+\hf}$ associated
with the links between sites:
\begin{equation}
\ell_{i} = \ee^{-(E_{i-\hf}+E_i)}, \qquad
r_{i} = \ee^{-(E_{i+\hf}+E_i)}
\label{equ:defE}
\end{equation} 
Note that site energies $E_i$ and transition
state energies $E_{i+\hf}$ are measured with respect to separate
arbitrary baselines, so they do not necessarily have to
be positive; but an interpretation
in terms of activated hopping processes becomes problematic
unless $E_i+E_{i\pm\hf}>0$.

Two subclasses of these general hopping models have been
studied quite extensively in the past~\cite{Bern79,
Alex81,BouGeo90,JS-dual,Machta85,MonBou96,Bertin03,Mon03}.
The first subclass is the (pure) trap model, which is the case
$E_{i+\hf}=0$ for all $i$, with the $E_i$ being
independent and identically distributed (i.i.d.); the
second is the (pure) barrier model, which has 
$E_i=0$ for all $i$, with the $E_{i+\hf}$ being i.i.d.
A duality between pure trap and pure barrier models
was discussed in~\cite{JS-dual}, which we now generalise.
The duality relation involves an inversion
of the potential, which swaps the meaning of transition state
energies and site energies: the various
definitions are illustrated in Fig.~\ref{fig:invert_all}.
To present this relation,
we define an alternative representation of the general hopping 
model, in which sites have indices $i+\hf$ (throughout this
article, $i$ is always an integer).
The hopping rate from $i-\hf$ to $i+\hf$ is $R_i$
and the rate for the reverse process is $L_i$.  
The master equation
for this process has a similar representation in terms
of state vectors $|i+\hf\rangle$, with a master
operator
\begin{equation}
\Whf = \sum_{i=1}^N  
\left( \Bigs|i+\hfs\Bigs\rangle - \Bigs|i-\hfs\Bigs\rangle \right) 
 \left( R_i\Bigs\langle i-\hfs\Bigs| - L_i\Bigs\langle i+\hfs\Bigs| \right).
\label{equ:Wb}
\end{equation}

If we choose $R_i=r_i$ and $L_i=\ell_{i+1}$, then $\Whf$ describes
the same model as $\Wint$, up to a simple relabelling of sites (site
$i$ in $\Wint$ is mapped to site $i-\hf$ in $\Whf$).
To describe instead
the inverted potential of Fig.~\ref{fig:invert_all}, we choose
\begin{equation}
R_i=\ee^{-(E_i+E_{i-\hf})}=\ell_i, \qquad L_i=
\ee^{-(E_i+E_{i+\hf})} = r_i
\label{equ:LRlr}
\end{equation}

With these definitions, the duality relation takes a simple
form: we define operators
\begin{eqnarray}
S 
 &=& \sum_{i=1}^{N} 
\left( |i+1\rangle - |i\rangle \right) \Bigs\langle i+\hfs\Bigs|
                 \ee^{-E_{i+\hf}}
\label{eq:S1st}
\\
&=& \sum_{i=1}^N 
|i\rangle \left( \ee^{-E_{i-\hf}}\Bigs\langle i-\hfs\Bigs|
 - \ee^{-E_{i+\hf}} \Bigs\langle i+\hfs\Bigs| \right) 
\end{eqnarray}
and 
\begin{eqnarray}
\Sbar &=& \sum_{i=1}^N 
\left( \Bigs|i+\hfs\Bigs\rangle - \Bigs|i-\hfs\Bigs\rangle \right)
 \langle i| \ee^{-E_i} 
\\
&=& \sum_{i=1}^N 
\Bigs|i+\hfs \Bigs\rangle \left( \langle i|\ee^{-E_i} -
\langle i+1| \ee^{-E_{i+1}} \right).
\label{eq:Sbar2nd}
\end{eqnarray}
Hence, by combining the first expression (\ref{eq:S1st}) for $S$ with
the second expression (\ref{eq:Sbar2nd}) for $\Sbar$, and \emph{vice versa},
one verifies that
\begin{equation}
\Wint = S \Sbar, \qquad \Whf = \Sbar S.
\label{equ:susy}
\end{equation}
It follows that
\begin{equation}
\Sbar \Wint = \Whf \Sbar,
\label{equ:dual_sbar}
\end{equation}
and that
\begin{equation}
\Wint S = S \Whf.
\label{equ:dual_s}
\end{equation}
These two relations express the key duality between the two master
operators from which all our other results are derived. They imply,
for example, that $\Wint$ and $\Whf$ have the
same spectrum of eigenvalues: if $|\psi\rangle$ is a right eigenvector
of $\Whf$ then $S|\psi\rangle$ is a right eigenvector
of $\Wint$ with the same eigenvalue. The exception is the singular case where
$S|\psi\rangle=0$, which can happen only when the eigenvalue is
zero. An analogous argument can be made in the other direction,
showing overall that the nonzero eigenvalues and associated eigenvectors of
$\Wint$ and $\Whf$ are in one-to-one correspondence with each other.

In fact, the structure of (\ref{equ:susy})
occurs in the context
of supersymmetric field theories 
(for a discussion of supersymmetry in the language of operators, 
applied to energy landscapes, see~\cite{tanase-nicola};
for a more general introduction, see~\cite{zinn-justin-susy}).
These hopping
models are very simple examples of supersymmetric
partners:
the spaces $\{ |i\rangle \}$
and $\{ |i+\hf\rangle \}$ can be interpreted
as zero- and one-fermion subspaces of a generalised
superspace.   Acting to the right,
the operator $S$ annihilates all
elements of $\{|i\rangle\}$ and $\Sbar$ annihilates
all elements of $\{|i+\hf\rangle\}$ so $S^2=\Sbar^2=0$:
these operators are the supercharges of the theory.  We
can then write $\Wint+\Whf=S\Sbar + \Sbar S$ which allows
us to identify the two models as supersymmetric partners.
\newcommand{\latt}{a}
\newcommand{\ann}{c}
\newcommand{\cre}{c^\dag}
\newcommand{\x}{x}
\newcommand{\Vhf}{V_{1/2}}
\newcommand{\Vint}{V_1}

It is also instructive to consider the continuum limit of our lattice
model.  Assume that the lattice spacing is $\latt$, 
so that the position of site $i$ is $\x_i=ia$.  
We define continuous functions $V_1(\x)$ and $V_{1/2}(\x)$
and take $E_i=V_1(\x_i)$ and $E_{i+\hf}=V_{1/2}(\x_{i+\hf})$.
Within the continuum limit, we represent the superspace 
explicitly, using a basis $|\x,n\rangle$, where $\x$ is the position
and $n=0,1$ distinguishes
the zero- and one-fermion subspaces.  That is, taking fermionic
operators $\ann,\cre$ with $\ann^2=(\cre)^2=0$ and $\ann\cre+
\cre\ann=1$, we take $\ann|\x,0\rangle=0$ and
$|\x,1\rangle=\cre|\x,0\rangle$.  To make contact with
the basis used for the lattice model, we identify 
$|i\rangle$ with $|\x_i,0\rangle$ and $|i+\hf\rangle$ with $|\x_i,1\rangle$.
If we then divide $S$ and $\Sbar$ by $a$ and
take the lattice spacing to zero, we arrive at
\begin{equation}
S = \ann \frac{d}{d\x}\ee^{-\Vhf(\x)}, \qquad
\Sbar = \cre \frac{d}{d\x}\ee^{-\Vint(\x)}
\end{equation}
The master (or Fokker-Planck) operators follow as 
\begin{eqnarray}
\Wint &=& S \Sbar = \ann\cre \frac{d}{d\x}\ee^{-\Vhf(\x)}
\frac{d}{d\x}\ee^{-\Vint(\x)},
\nonumber \\ 
\Whf &=& \Sbar S = \cre\ann \frac{d}{d\x}\ee^{-\Vint(\x)}
\frac{d}{d\x}\ee^{-\Vhf(\x)}
\end{eqnarray}
This makes it obvious that the duality just swaps the trap and barrier
parts of the potential, i.e.\ inverts the energy landscape. In the
case without thermal activation, which corresponds to $\Vhf=-\Vint=V$,
the duality reduces to the standard one for diffusion in the
potentials $V$ and $-V$.
One part 
of this duality was used
recently in~\cite{JPA-Tailleur} to map boundary-driven steady states
with current to current-free equilibrium states.  
Briefly,
if $P(\x)$ is a steady state probability distribution
of $\Wint$ in the presence of a boundary field so that
$\Wint|P(\x),0\rangle = 0$, then
$|\tilde{P}(\x),1\rangle=\Sbar|P(\x),0\rangle$ obeys 
$S|\tilde{P}(\x),1\rangle=0$.  Thus, $|\tilde{P}(\x),1\rangle$ is
 a steady state of $\Whf$, but the current in this state 
is $-\ee^{V}(d/d\x)\ee^{-V}|\tilde{P}(\x),1\rangle=
-\ee^{V}\cre S|\tilde{P}(\x),1\rangle=0$. The same approach also works on
the lattice, and indeed the rate transformation (182) in
Ref.~\cite{JPA-Tailleur} is the same as our (\ref{equ:LRlr}).

The total master operator combining the dynamics in the zero and
one-fermion spaces can be made Hermitian by a standard similarity
transformation: with $X=\ann\cre \ee^{\hf \Vint(\x)} + \cre\ann
\,\ee^{\hf \Vhf(\x)}$ one has
\begin{eqnarray}
\fl H&=&X^{-1}(\Wint+\Whf)X \nonumber \\
\fl &=& \ann\cre \ee^{-\hf\Vint(\x)}\frac{d}{d\x}\ee^{-\Vhf(\x)}
\frac{d}{d\x}\ee^{-\hf\Vint(\x)} 
+ \cre\ann\, \ee^{-\hf\Vhf(\x)}\frac{d}{d\x}\ee^{-\Vint(\x)}
\frac{d}{d\x}\ee^{-\hf\Vhf(\x)}
\end{eqnarray}
Using $\ann\cre=1-\cre\ann$ to extract the extra contribution from the
one-fermion subspace, one can simplify this to
\begin{eqnarray}
  \fl H & = & \ee^{-\hf\Vint(\x)}\frac{d}{d\x}\ee^{-\Vhf(\x)}
  \frac{d}{d\x}\ee^{-\hf\Vint(\x)}  
\nonumber \\ \fl & &
  - \cre\ann\, \ee^{-\Vint(\x)-\Vhf(\x)}
  \left[\frac{1}{4}(\Vint'(\x)^2-\Vhf'(\x)^2)-\hf(\Vint''(\x)-\Vhf''(\x))\right]
\end{eqnarray}
For the 
`standard' case with uniform diffusion constant 
$\Vhf=-\Vint=V$, the one-fermion term
simplifies to $-\cre\ann V''(\x)$ and this is exactly the term that
was used e.g.\ in~\cite{tanase-nicola,zinn-justin-susy} 
to construct dynamics that
(in one dimension, and in the one-fermion subspace) converges to
maxima rather than minima of the potential.

\subsection{Choice of boundary conditions}
\label{sec:bc}

In the previous section, we discussed systems with detailed balance
and periodic boundaries.  We now discuss how the duality relation
applies on finite chains with reflecting or absorbing
boundaries.  We note in passing that some of these
relations may be generalised both to boundary-driven models
and to those with a finite bias acting in the bulk~\cite{forthcoming}.  
However, 
for this work, we restrict ourselves to models without a
steady-state current.

In the case of periodic boundary conditions, the derivation of our
duality relations required that the rates
satisfy the global constraint $\prod_i \ell_i = \prod_i r_i$, in order to
guarantee detailed balance. An
alternative that avoids this constraint is to use finite chains with reflecting
or absorbing boundary conditions.  These can be obtained
by allowing zero rates in the periodic system.  For a
system with reflecting boundaries, we take $\ell_1=r_N=0$ in $\Wint$.
In the notation of energies, we set formally $\ee^{-E_\hf}=0$:
sites $1$ and
$N$ are then reflecting boundaries because they are separated by an
infinite barrier. This has no effect on the equilibrium steady state, which now
satisfies detailed balance whatever our choice for the remaining nonzero rates. 
The duality transformation carries through as before, 
resulting in a system with $L_N=R_1=0$. 
In this system there are no
transitions out of site $\hf$, which is therefore absorbing.
The duality thus relates models with reflecting and absorbing
boundaries, and as before these models share the same eigenvalues and
their
eigenvectors (for nonzero eigenvalues) are related through the
operators $S$ and $\Sbar$. Note that for $\Whf$, the steady state that
is reached in the long-time limit has the particle fully localised on
the absorbing site $\hf$.
Detailed balance still holds, with e.g. the propagators 
obeying~(\ref{equ:detbal}) below; 
the balance of transitions in the steady state is trivial 
because there are no transitions taking place at all.

One can go further by distinguishing whether
the absorbing site in $\Whf$ is reached from the left (from site
$\frac32$) or from the right (from site $N-\hf$), and accordingly split the
absorbing site into two sites $\hf$ and $N+\hf$. The duality
relation to the system $\Wint$ with two reflecting boundaries
then holds as before. The only difference is 
that, because the system now has two absorbing sites, the 
master operator has two zero eigenvalues. The two corresponding right
eigenvectors are localised on the absorbing sites, while the
elements of the left eigenvectors give the probabilities that
a particle initially on a given site will end up on 
either one 
of the absorbing sites (this will be discussed further
in later sections).

Finally, one may also take
an operator $\Wint$ with one reflecting boundary
and an absorbing site at the other boundary: this can be done by
setting $\ee^{-E_\hf}=0$ and $\ee^{-E_N}=0$, so that $\ell_1=\ell_N=r_N=0$.
In the dual model of
this system, $R_1=R_N=L_N=0$, so that the reflecting boundary at site
1 maps to an absorbing boundary at 
site $\hf$ and \emph{vice versa} for sites $N$ and $N-\hf$.

\subsection{Alternative formulation, and generalised duality}

So far, our results apply to models in which the steady state has
zero current (that is, periodic chains with global detailed balance,
and chains with reflecting or absorbing boundaries).  However,
there is a slightly modified duality relation that holds even
for periodic chains without detailed balance, where
the steady state has finite current.  Instead of factorising
$\Wint = S \Sbar$ as in (\ref{equ:susy}), one may instead
write
\begin{equation}
\Wint = - D J
\end{equation}
with $D=\sum_i (|i\rangle - |i+1\rangle) \langle i+\hfs|$
and $J=\sum_i |i+\hfs\rangle (r_{i} \langle i| - \ell_{i+1} \langle i+1|)$.
[In the continuum limit, one has $\Wint = -\frac{d}{dx} J$ where
$J=-\ee^{-\Vhf(\x)} \frac{d}{d\x} \ee^{-V_1(\x)}$ 
is the probability current operator, so the master (Fokker-Planck)
equation has the form $\partial_t P=-\nabla (JP)$ 
as usual, where $JP$ is the probability current, which
is linear in $P$. 

With these definitions, one may verify that
\begin{equation}
\Whf = - (J D)^\dag
\label{equ:JDdag}
\end{equation}
or equivalently that 
\begin{equation}
\Wint D = -DJD = D (\Whf)^\dag.
\label{equ:dual_D}
\end{equation}
If detailed
balance holds, the relation $\langle i+\hfs | \Whf | j+\hfs\rangle \ee^{E_{j+\hf}}=
\langle j+\hfs | \Whf |i+\hfs \rangle \ee^{E_{i+\hf}}$ may be combined with
(\ref{equ:JDdag}) to recover (\ref{equ:dual_s}).
(One way is to define
$E=\sum_i |i+\hf\rangle\langle i+\hf|\ee^{E_{i+\hf}}$ and to note that
$D=-SE$, $J=E^{-1}\Sbar$, and by detailed balance $\Whf=E(\Whf)^\dag
E^{-1}=-EJDE^{-1}=\Sbar S$.)

We also note that
the operator $D$ may not be inverted, since $D\sum_i |i+\hf\rangle=0$.
However, if one considers for example a periodic chain of $N$ sites
with an absorbing site 1 and a reflecting barrier $\hf$, so that
$r_1=\ell_1=r_N=0$, then one may
write $\Wint = -D' J$ and $\Whf = -(JD')^\dag$ 
with $D'=D+|1\rangle\langle \hf|$.  In that
case, it may be verified that
the inverse of $D'$ does exist, and the duality transformation
becomes
\begin{equation}
(D')^{-1} \Wint D' = (\Whf)^\dag
\end{equation}

We note that the transformation operators $D$ and $D'$ do not depend
on the disordered rates $r_i$ and $\ell_i$.  An invertible
transformation that is independent of disorder was used by
Sch\"utz and Mussawisade~\cite{schutz} to study a reaction-diffusion model
in a similar disordered environment.  The duality transformation
(enantiodromy) ${\cal D}$
used in~\cite{schutz} is related to the transformation $D'$,
and can be used to prove some 
of the relations
for propagators that we discuss in the next section.  However, analysis
of the reaction-diffusion model is much more complex than the single
particle system considered here.  In the following,
we restrict our analysis to the single-particle
case, explaining which results may be obtained by alternative
means.

\section{Propagators}
\label{sec:prop}

We now consider the propagators of these hopping models.
For convenience, we first consider periodic boundaries in 
the case where detailed balance holds.  Generalisations
to reflecting/absorbing boundaries are straightforward using the
approach discussed above (Sec.~\ref{sec:bc}).
For the model $\Wint$, we define the propagator $\Gint_{n,m}(t)$ as
the probability that the particle is on site $n$, given that it
was on site $m$ a time $t$ earlier: 
\begin{equation}
\Gint_{n,m}(t) = \langle n | \ee^{\Wint t} | m\rangle
\end{equation}
Similarly, for the model $\Whf$, we define
\begin{equation}
\Ghf_{n+\hf,m+\hf}(t) = \Bigs\langle n+\hfs \Bigs| \ee^{\Whf t} \Bigs| m+\hfs\Bigs\rangle
\end{equation}
\begin{figure*}
\hfill \epsfig{file=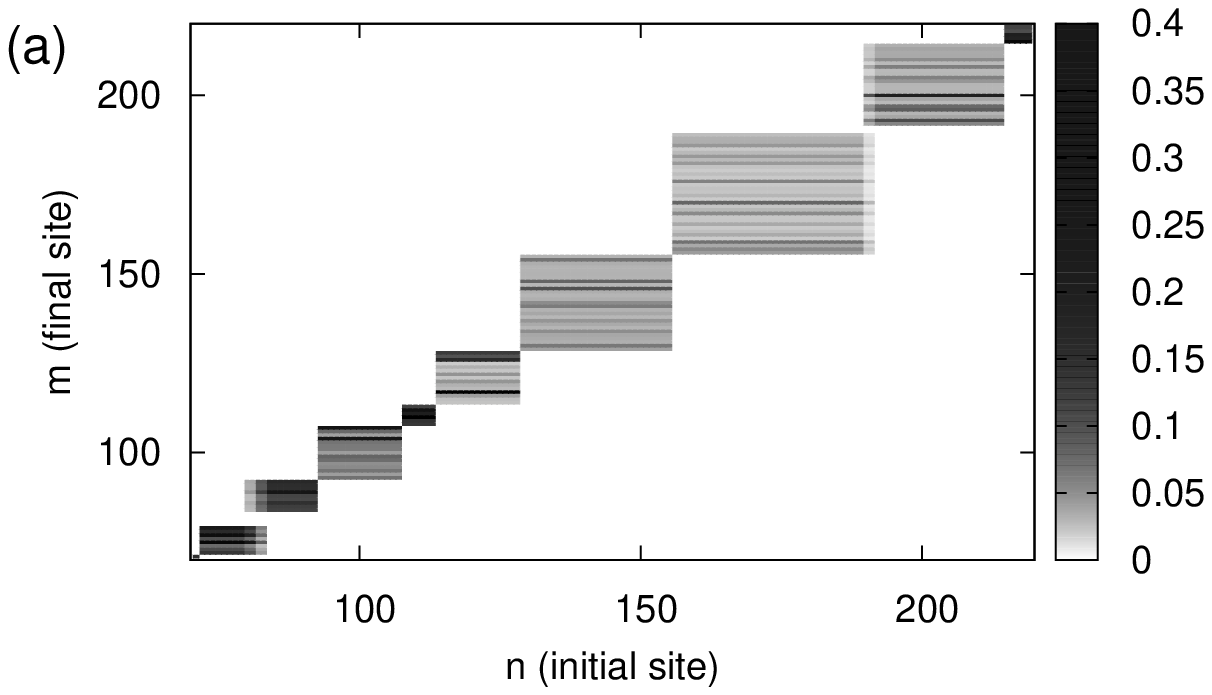,width=0.48\textwidth}
\epsfig{file=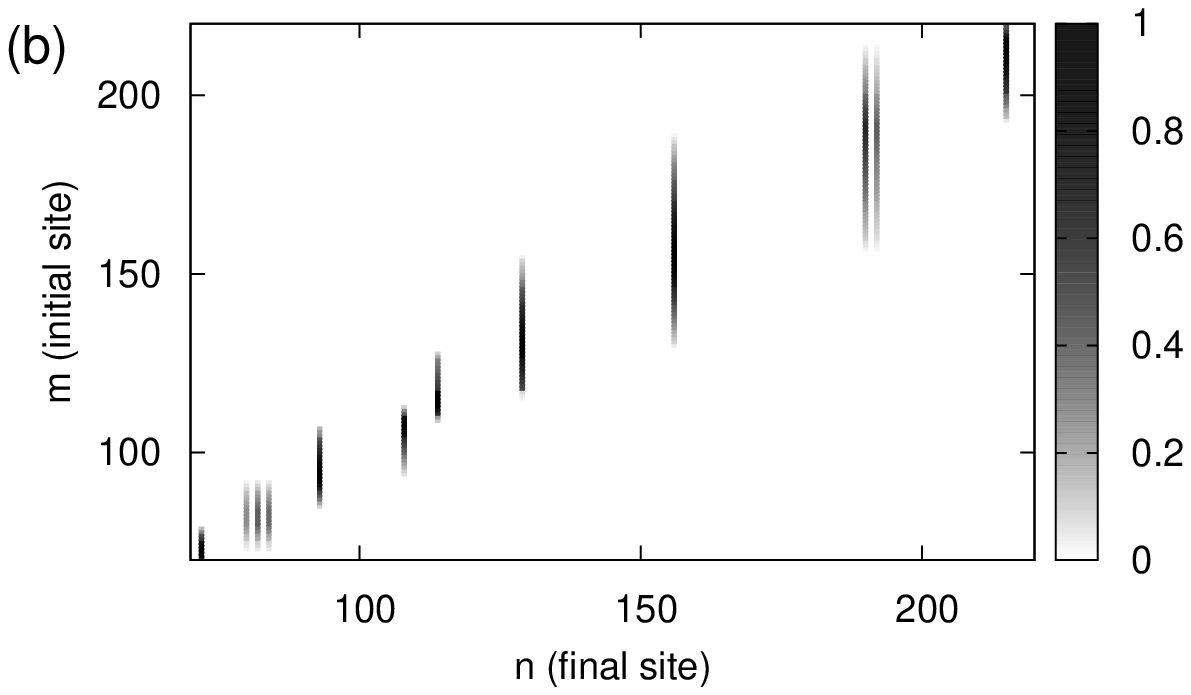,width=0.48\textwidth}
\caption{Plots of Green's functions that illustrate the duality
relation between the propagators.  The plots were generated using the effective
dynamics scheme described in section~\ref{sec:eff}, but
(\ref{equ:dual_prop}) applies to propagation under the
effective dynamics as well as to propagation with real dynamics.
(Left) Propagator $G_{m,n}$, describing motion from site $n$ to site $m$
in a model of mixed trap-barrier
type with $\mub=3$ and $\mut=0.5$ (see 
Sec.~\ref{sec:mixed} for definitions). The `squares' represent 
`trap regions' within which the particle is locally equilibrated.  The
structure within the trap depends on the final site
energies.  These energies are randomly distributed, so that sites with low
energy $-E_i$ give rise to dark `stripes'. However, local
equilibration within the trap means that the propagator
depends only weakly on the initial site $n$ for all the $n$
within a single trap.  (Right) Propagator $G_{n,m}$ 
in the dual model of that shown on the left: note
that $m$ is now the initial site and $n$ the final site.  The
probability for the final position is localised  on a few deep
`trap' sites.  Eq.~(\ref{equ:dual_prop}) states that for given values of $m$ and $n$,
the gradient of the right plot
in the vertical ($m$)-direction is equal to the 
gradient of the left plot in the horizontal ($n$)-direction.  }
\label{fig:dual_illust}
\end{figure*}
Detailed balance holds for both models so we have, bearing in mind our
sign convention for energies, $\Gint_{n,m}(t)\ee^{E_m}  =
\Gint_{m,n}(t)\ee^{E_n}$ or  
\begin{equation}
\ee^{-E_n} \Gint_{n,m}(t) = \ee^{-E_m} \Gint_{m,n}(t) ,
\label{equ:detbal}
\end{equation}
with a similar relation for $\Ghf$. We now appeal to the duality relation
$\Sbar \Wint = \Whf\Sbar$ to obtain
$\langle n+\hf |\Sbar \ee^{\Wint t} | m \rangle 
=\langle n+\hf | \ee^{\Whf t} \Sbar | m \rangle$.  Using
the explicit form of $\Sbar$, we arrive at
\begin{equation}
\fl \ee^{-E_n}\Gint_{n,m}(t) -  \ee^{-E_{n+1}}\Gint_{n+1,m}(t)
 =
\left[ \Ghf_{n+\hf,m+\hf}(t) - \Ghf_{n+\hf,m-\hf}(t) \right] \ee^{-E_m},
\label{equ:gsbar}
\end{equation}
from which detailed balance for $\Wint$  implies
\begin{equation}
\Gint_{m,n}(t) - \Gint_{m,n+1}(t)
  = 
\Ghf_{n+\hf,m+\hf}(t) - \Ghf_{n+\hf,m-\hf}(t) .
\label{equ:dual_prop}
\end{equation}
This result generalises Eq.~(5) of Ref.~\cite{JS-dual}.  
Equation~(\ref{equ:dual_prop}) may also be proved directly
using the relation~(\ref{equ:dual_D}).  Hence, this relation
between propagators applies on periodic chains, regardless
of whether detailed balance holds.
It holds for all times, including cases where the limit of large
time is taken at fixed system size.  

Two comments are in order here.  Firstly,
a relation similar to~(\ref{equ:dual_prop}) was proven
for propagators in the reaction-diffusion model of~\cite{schutz}.
That model has pair annihilation, so we consider an 
initial state with two well-separated particles.  If one then
takes the limit $N\to\infty$ at fixed time, one arrives at 
two independently propagating particles 
which can be shown to satisfy~(\ref{equ:dual_prop}).  
However, the mapping of~\cite{schutz}
cannot be used to prove (\ref{equ:dual_prop}) for 
large times and periodic chains of
fixed length: in that case, it would be natural to use an initial state
with exactly one particle but the mapping then breaks down
(except for specific cases where the periodic chain
is broken into segments by absorbing sites and reflecting barriers).
Secondly, for the specific case where the $E_i$ are chosen freely but
$E_{i+\hf}=0$ for all $i$,
the relation~(\ref{equ:dual_prop}) was given in Ref.~\cite{JS-dual},
but it is also implicit in~Ref.~\cite{DenteneerErnst}. 
(In the notation there, it reads  $f(q)
\hat\Gamma^{\mathrm{B}}_{qq'}(z) = \hat\Gamma^{\mathrm{T}}_{qq'}(z) f(q')$ 
and can be derived by inserting Eqs.~(2.15) and (2.16) into (2.19)
of Ref.~\cite{DenteneerErnst} and expanding appropriately.)

The equality~(\ref{equ:dual_prop}) relates differences
with respect to the initial site of propagators in the two hopping
models.  If the propagator for one model is known, the other
can be calculated by successive application of this equation.  
An illustration of this relation is shown in 
Fig.~\ref{fig:dual_illust}.  
We also note that (\ref{equ:dual_prop}) is symmetric between the two 
master operators: any set of transition rates can be interpreted
either as a model of type $\Wint$ or as a model of type $\Whf$.  
To reinforce this symmetry, we note that the 
relation 
$\langle n|S\ee^{\Whf t}|m-\hf\rangle= \langle n|\ee^{\Wint t} S|m-\hf\rangle$
implies that 
\begin{equation}
\fl   \ee^{-E_{n-\hf}} \Ghf_{n-\hf,m-\hf}(t)
-  \ee^{-E_{n+\hf}} \Ghf_{n+\hf,m-\hf}(t)
= \left[ \Gint_{n,m}(t) - \Gint_{n,m-1}(t)\right] \ee^{-E_{m-\hf}}
\label{equ:gs}
\end{equation}
which is the same relation as (\ref{equ:gsbar}), but with the
original model $\Wint$ expressed in the form $\Whf$ and 
\emph{vice versa}.

The simple form of (\ref{equ:dual_prop}) means that
we can take the disorder average. As long as the disorder is
translationally invariant, e.g.\ if the transition state
energies $\{E_{i+\hf}\}$ and site energies $\{E_i\}$ are taken
from two (possibly different) translationally invariant distributions, the
disorder-averaged propagators $\overline{G}_{n,m}(t)$ 
will depend only on $k=n-m$.  
Defining then $\Delta_k=\Gbarint_k(t) -
\Gbarhf_{-k}(t)$, the relation (\ref{equ:dual_prop}) becomes
$\Delta_{-k}=\Delta_{-k-1}$. 
Thus, $\Delta_k$ is a constant; but
$\sum_{k}\Gbarint_k(t)=1$ and similarly for $\Gbarhf_k(t)$,
so $\sum_k \Delta_k=0$ and the constant has to vanish. 
One concludes that $\Gbarint_{k}(t)
= \Gbarhf_{-k}(t)$. Re-instating the notation with separate
initial and finite site labels, we have
\begin{equation}
\Gbarint_{m,n}(t) 
  = 
\Gbarhf_{n,m}(t)
  = 
\Gbarhf_{m,n}(t)
\end{equation}
where the last equality follows from left-right symmetry. Remarkably,
then, the disorder-averaged propagators of the dual models 
are equal on all scales of length and time. Again, this generalises
our earlier statement~\cite{JS-dual} relating average propagators in pure
trap and barrier models.
A similar result applies for disorder distributions which break
left-right symmetry, as long as they remain translationally
invariant.  Take for example a model described by $\Wint$
with site and barrier energies from a translationally
invariant distribution, and then
modify the rates according to a site-independent prescription,
so that the left-going and right-going rates acquire different
distributions.  With periodic
boundary conditions, translation invariance again holds and
we obtain by the same arguments as above
\begin{equation}
{\Gbarint_{m,n}}(t) 
 = 
{\Gbarhf_{n,m}}(t),
\label{equ:gbar_dir}
\end{equation}
with both sides again depending only on the difference $k=n-m$. The
same result applies on long chains with reflecting boundaries, as long
as we are far away from these boundaries so that translation
invariance is not broken.

\section{Effective dynamics scheme}
\label{sec:eff} 

Several effective dynamics and renormalisation schemes have been proposed 
to approximate propagators for motion in random 
potentials~\cite{DMF,Machta85,Mon03,Garel-Monthus}.  The most
notable success in this area is the work of le Doussal,
Monthus and Fisher~\cite{DMF}, which we refer to as DMF.
They considered the Sinai model~\cite{sinai}, in which the $r_i$ and $\ell_i$
are independently and identically distributed, so that the site
energies $E_i$ follow a random walk in real space.  For long
time scales in that model, DMF found a renormalisation group (RG) scheme
that can be implemented analytically and gives exact
results for a variety of observables.  

Later, Monthus~\cite{Mon03} 
applied a related scheme to the pure trap model, in 
which $\ell_i=r_i$ and these rates are independently
and identically distributed with a power-law distribution
$P(r_i)\propto r_i^{(1/\mu)-1}$.
We refer to this scheme as Mon03: it can be
treated analytically and its predictions are exact 
in the limit of large $\mu$ (our notation follows that of
\cite{JS-dual}: to arrive at the notation of~\cite{Mon03}, 
replace $\mu$ by $\frac1\mu$).  
More recently, Monthus and Garel (MG) derived a general RG
scheme~\cite{Garel-Monthus} 
that is not restricted to one dimension
nor to single-particle systems.  Like the schemes of Refs.~\cite{DMF,Mon03},
the MG scheme is effective when disorder is the dominant source of fluctuations.  
Finally, we recently~\cite{JS-dual}
introduced a modified scheme for pure trap and barrier
models that respects the duality symmetry~(\ref{equ:dual_prop})
but requires a computational implementation.  
This prescription accounts for effects that were neglected
in previous schemes, and reduces to the
Mon03 scheme in the limits in which that method is exact.

Here, we generalise the scheme of~\cite{JS-dual} to general disordered
potentials (in one dimension).  We arrive at a method that encompasses
the DMF and Mon03 methods in the limits when they are exact,
and is also consistent with the duality relation~(\ref{equ:dual_prop}).
Our method also shares features with that of MG, but it respects the duality
symmetries discussed above, while the MG scheme does not.
We describe the application of this scheme to models parameterised
in the form of
$\Wint$, with site energies $E_i$ and transition state
energies $E_{i+\hf}$.  For notational convenience, we simply
denote these master operators by $W$ from now on.

\subsection{Definition of the effective dynamics scheme}
\label{sec:rg}

The idea of the effective dynamics scheme is to describe motion on long
time scales in terms of a coarse-grained set of co-ordinates
in which the system evolves slowly.  We have developed
two versions of the scheme: in the main text we describe
the more physical and intuitive version that also gives the most
accurate description of motion in the system.  We discuss
the justifications of our scheme in later sections (\ref{app:eff}
and Sec.~\ref{sec:numerics}): in particular,
the second version of the scheme can be justified more formally,
and it gives similar results, but with larger errors for finite $\mu$.

The scheme is parameterised by a time scale $\Gamma^{-1}$.  
On that time scale, we partition the 1d chain into 
effective trap and effective barrier regions.  The coarse-grained
slow co-ordinates are the occupancies of the effective
traps:
\begin{equation}
p_\alpha(t) = \sum_{i=a_\alpha}^{b_\alpha} p_i(t).
\label{equ:p_slow}
\end{equation}
We continue to use Roman indices $i,j,\dots$ for the original 
site labels of the model, while Greek indices $\alpha,\beta,\dots$
label the slow co-ordinates.  Thus, $a_\alpha$
and $b_\alpha$ are the leftmost and rightmost sites 
within effective trap $\alpha$.  The traps form an ordered set along the
chain, so we have $b_{\alpha-1} < a_\alpha \leq b_\alpha < a_{\alpha+1}$.
We refer to the regions between these effective traps as effective
barriers.

For a given value of $\Gamma$, the effective dynamics scheme
gives (i) an approximation for the propagator 
$G_{mn}(\Gamma^{-1})$ and (ii) an approximate equation
of motion for the slow co-ordinates, valid for times $t>\Gamma^{-1}$.
In addition, the scheme specifies (iii) how the set of
effective traps evolves as $\Gamma$ is decreased towards zero.

Physically, the idea is that motion within effective traps
is fast, while motion between traps is slow.  Assuming
equilibration within effective traps,
we have for sites within
an effective trap region,
\begin{equation}
G_{ji}(\Gamma^{-1}) \simeq \ee^{E_j-F_\alpha},
\qquad a_\alpha \leq i,j \leq b_\alpha
\label{equ:prop_p1}
\end{equation}
where we identify the free energy of effective trap $\alpha$:
\begin{equation}
\ee^{F_\alpha} = \sum_{i=a_\alpha}^{b_\alpha} \ee^{E_i}
\label{equ:Fa}
\end{equation}
If site $i$ is between traps $\alpha-1$ and $\alpha$, that is
$b_{\alpha-1} < i < a_\alpha$, then 
it may relax into either trap before equilibrating there, and
we have 
\begin{equation}
G_{ji}(\Gamma^{-1}) \simeq \left\{
\begin{array}{ll} (1-v_i^{(\alpha-\hf)}) \ee^{E_j-F_{\alpha-1}}, 
  &  a_{\alpha-1} \leq j \leq b_{\alpha-1}
  \\
                v_i^{(\alpha-\hf)} \ee^{E_j-F_{\alpha}}, 
  &  a_{\alpha} \leq j \leq b_{\alpha}
\end{array} \right.
\label{equ:prop_p2}
\end{equation}
where $v_i^{(\alpha-\hf)}$ is the probability that a particle
initially on a site $i$ between traps $\alpha-1$ and $\alpha$ 
relaxes first into trap $\alpha$.
These probabilities can be obtained by considering propagation
on a chain with absorbing sites $b_{\alpha-1}$ and $a_\alpha$,
and calculating the probability of absorption into each of these
sites for a given initial site $i$.  The result is
\begin{equation}
v_i^{(\alpha-\hf)}=\ee^{-F_{\alpha-\hf}} \sum_{j=b_{\alpha-1}}^{i-1} 
\ee^{E_{j+\hf}} 
\end{equation}
with 
\begin{equation}
\ee^{F_{\alpha-\hf}}=\sum_{j=b_{\alpha-1}}^{a_{\alpha}-1} \ee^{E_{j+\hf}}.
\label{equ:Fahf}
\end{equation}
This completes point (i) above.  Turning to point (ii), the equations
of motion for the slow degrees of freedom are
\begin{equation}
\partial_t p_\alpha(t) =  \ell_{\alpha+1} p_{\alpha+1}(t)
 + r_{\alpha-1} p_{\alpha-1}(t) - (\ell_\alpha + r_\alpha) p_\alpha(t)
\label{equ:master_slow}
\end{equation}
with
\begin{equation}
r_\alpha = \ee^{-F_\alpha-F_{\alpha+\hf}}, \qquad
\ell_\alpha = \ee^{-F_\alpha-F_{\alpha-\hf}}
\label{equ:rla}
\end{equation}
That is, hopping takes place only between nearest neighbours,
and the rates are given by the free energies of the
effective traps $F_\alpha$ and `transition state
free energies' $F_{\alpha+\hf}$.

It remains to discuss point (iii) above: how the set of effective traps
depends on the time scale $\Gamma$.  
Initially each effective trap 
contains a
single site $i$, and each effective barrier a single transition site
$i+\hf$.
Each stage of our effective dynamics begins by
calculating rates for motion from trap $\alpha$:
we calculate a renormalised rate $\rho_\alpha$
associated with the rate $r_\alpha$, i.e.\ with leaving the trap to
the right:
\begin{equation}
\fl \rho_\alpha = \hf(r_\alpha + \ell_\alpha + \ell_{\alpha+1})
+ \hf \sqrt{ r_\alpha^2 + 2 r_\alpha ( \ell_\alpha + \ell_{\alpha+1} )
 + (\ell_\alpha-\ell_{\alpha+1})^2 }.
\label{equ:lam2}
\end{equation}
and similarly for motion to the left
\begin{equation}
\fl \lambda_\alpha = \hf(\ell_\alpha + r_\alpha + r_{\alpha-1})
+ \hf \sqrt{ \ell_\alpha^2 + 2\ell_\alpha ( r_\alpha + r_{\alpha-1} )
 + (r_\alpha-r_{\alpha-1})^2 },
\label{equ:lam1}
\end{equation}
Notice that $\rho_\alpha$ is symmetric in $\ell_\alpha$ and
$\ell_{\alpha+1}$. In a (renormalised) landscape similar to that for
pure traps, $\ell_\alpha$ would be comparable to $r_\alpha$ and so
$\rho_\alpha\approx r_\alpha+\ell_\alpha$ if the other rate
($\ell_{\alpha+1}$) is small. Conversely, in a landscape resembling
pure barriers, one would have $\ell_{\alpha+1}$ of the same order as
$r_\alpha$ and so to leading order
$\rho_\alpha \approx r_\alpha+\ell_{\alpha+1}$.  
The expressions
(\ref{equ:lam2},\ref{equ:lam1}) cover both of these limits but extend them to
general landscapes.  

Having calculated the $\rho_\alpha$ and $\lambda_\alpha$, we select
the largest of these rates across all traps, and update
$\Gamma$ to this largest rate.  
At this point, several cases arise, which are 
illustrated in the following section,
with reference to the example landscape in Fig.~\ref{fig:rg_new}.
Here we give the rules: Supposing that the
largest rate is $\rho_\alpha$, we now remove trap $\alpha$.
We make a case distinction, depending on
whether the local landscape is nearer to the pure trap or the pure
barrier case, as in the discussion of the rate-dependence of
$\rho_\alpha$ above.  If $\ell_{\alpha+1}>\ell_\alpha$ we 
combine trap $\alpha$ into trap $\alpha+1$, leading to a
new effective trap containing sites $a_\alpha\dots
b_{\alpha+1}$. Barrier region $\alpha+\hf$ is removed.
Conversely, if
$\ell_{\alpha+1}<\ell_\alpha$, we remove trap $\alpha$
from the list of slow co-ordinates:
this amounts to combining barrier regions
$\alpha\pm\hf$. 
For the case where the largest renormalised rate is $\lambda_\alpha$,
associated with motion to the left, the rules are 
similar, in accordance with left-right symmetry: if
$r_{\alpha-1}>r_\alpha$ we combine traps $\alpha-1$ and $\alpha$;
otherwise we remove trap $\alpha$
which amounts to combining the two barrier regions $\alpha\pm\hf$.

Having combined the appropriate
traps or barriers, we finally
recalculate the free energies $F_\alpha$
and $F_{\alpha+\hf}$ that are affected by the change,
and hence obtain new hopping rates
$r_\alpha$ and $\ell_\alpha$, and new renormalised rates
$\rho_\alpha$ and $\lambda_\alpha$.
From here on the
process is iterated, i.e.\ we find the largest rate among the
$\{\rho_\alpha,\lambda_\alpha\}$, update $\Gamma$, and merge the
appropriate traps or barriers. 

For any given $\Gamma$, we can therefore calculate the approximate
propagator and the equation of motion for the slow degrees of freedom.
We emphasise that
while we have defined rates $r_\alpha,\ell_\alpha,\lambda_\alpha,
\rho_\alpha$, free energies $F_\alpha,F_{\alpha+\hf}$ and parameters
$a_\alpha,b_\alpha,v_i^{(\alpha-\hf)}$, all of these quantities are
fixed if we specify the set of slow co-ordinates (parameterised
in terms of the $a_\alpha$ and $b_\alpha$), together with the full
set of bare energies $E_i,E_{i+\hf}$.  We discuss below how the progress
of the scheme can be written in terms of a $\Gamma$-dependent
projection operator, using the notation of Sec.~\ref{sec:dual}.
However, we first give an illustrative example of the effective
dynamics in action.

\subsection{Example of effective dynamics and comparison with DMF scheme}
\label{sec:example_rg}

\begin{figure*}
\hfill \includegraphics[width=0.8\textwidth]{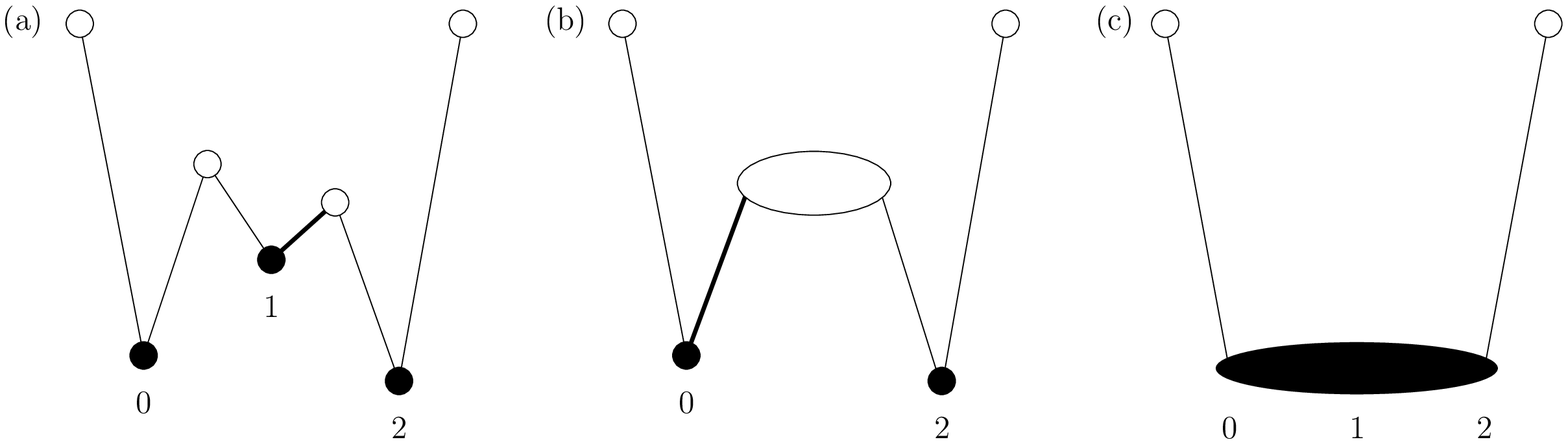}
\caption{Illustration of the RG scheme introduced
in this paper, discussed in the main
text and in~\ref{app:eff}.  
As in Fig.~\ref{fig:invert_all}, sites are
shown with closed circles and transition states
with open circles.  However, we now introduce effective
trap regions, represented by filled ovals, and
effective barrier regions, 
represented by open ovals.
(a)~Energy landscape for a hopping model with rates
satisfying~(\ref{equ:eg_rates}).
The fastest rate $r_1$ is associated with the energy
difference $E_1+E_{\frac32}$: the link associated with
this rate is emphasised by a bold line.  (b)~Energy landscape
obtained after one stage of renormalisation, with
slow degrees of freedom $\{\dots,p_0(t),p_2(t),\dots\}$.
The transition
states $\frac12$ and $\frac32$ and site $1$ have
been incorporated into a composite transition state,
which we refer to as an effective barrier region.
Within the effective dynamics, a particle
initially on site $1$ relaxes either to site 0 or site 2,
according to~(\ref{equ:prop_p2}).
The link associated with the largest rate 
in the renormalised master operator is again emphasised.
(c)~Energy landscape after the second stage of renormalisation,
with slow degrees of freedom $\{\dots,p_0(t)+p_1(t)+p_2(t),\dots\}$
The two effective traps on sites $0$ and $2$ have been
combined into a single effective trap.  Within the
effective dynamics, particles initially
on any of these sites relax to an equilibrium distribution
over sites $0,1,2$.
}
\label{fig:rg_new}
\end{figure*}

We illustrate the effective dynamics
using the example landscape shown in Fig.~\ref{fig:rg_new}, concentrating
on sites $0,1,2$ of a long chain.  
Consistent with Fig.~\ref{fig:rg_new}a, we choose the rates
to lie in three well-separated sectors: 
\begin{equation}
r_1, \ell_1 \gg
r_0,\ell_2 \gg \ell_0, r_2.
\label{equ:eg_rates}
\end{equation}  
For concreteness we also
take $r_1>\ell_1$ (i.e.\ $E_{\hf}>E_{\frac32}$), $\ell_2>r_0$ and
$E_2>E_0$.

At the initial stage of the dynamics, sites $0,1,2$ each
constitute an effective trap: the effective equation of motion 
(\ref{equ:master_slow})
coincides with the original master equation (\ref{equ:master}) and the
relevant degrees of freedom are simply the original site occupancies
$\{\dots,p_0(t),p_1(t),p_2(t),\dots\}$
where the $(\dots)$ indicate that we are concentrating
on part of a large chain.  Taking the original rates
$r_i$ and $\ell_i$, we
evaluate the parameters $\rho_\alpha$ and $\lambda_\alpha$, and the largest
of these is $\rho_1=r_1+\ell_1+\ell_2+\hf\sqrt{r_1^2+2r_1(\ell_1+\ell_2)
+(\ell_1-\ell_2)^2}$.
(To leading order in the largest rates $r_1$ and $\ell_1$,
$\rho_1=\lambda_1=r_1+\ell_1$; but the correction from $\ell_2$ can then
be shown to make $\rho_1>\lambda_1$.)
Thus, the first step of the effective dynamics occurs as
$\Gamma$ is decreased through $\rho_1$.
From (\ref{equ:eg_rates}), we have
$\ell_2<\ell_1$, so the rules state that we remove the trap
on site 1.  Physically, the idea is that the transition
state energies $E_\hf$ and $E_{\frac32}$ are similar to each other,
so they are combined into an effective barrier. On the
other hand, the particle spends very little time on site $1$,
compared to sites $0$ and $2$, so the co-ordinate $p_1(t)$
relaxes quickly to a small value and is no longer a relevant
(slow) co-ordinate.

The remaining slow co-ordinates are therefore
$\{\dots,p_0(t),p_2(t),\dots\}$, and the energy
landscape on this time scale is shown in Fig.~\ref{fig:rg_new}b.  
From~(\ref{equ:master_slow}),
the equations of motion for these two co-ordinates at this stage
are
\begin{eqnarray}
\partial_t p_0(t) &=& r_{-1} p_{-1}(t) + \ell_2' p_2(t) -
(r_0' + \ell_0) p_0(t) 
\nonumber \\
\partial_t p_2(t) &=& r_{0}' p_{0}(t) + \ell_3 p_3(t) -
(r_2 + \ell_2') p_2(t) 
\label{equ:rg_int_eg}
\end{eqnarray}
where the unprimed rates $r_i$ and $\ell_i$ are the original
hopping rates among the sites of the model, but
two new rates have appeared:
$r_0'=\ee^{-E_0-F}$ and $\ell_2'=\ee^{-E_2-F}$.  
Here $\ee^{F} \equiv  \ee^{E_\hf} + \ee^{E_{\frac32}}$, and $F$ 
is the `transition state free
energy' for the effective barrier region between sites $0$ and $2$,
constructed according to (\ref{equ:Fahf}).  
One may also construct the effective propagator
on the time scale $\rho_1^{-1}$ in accordance with (\ref{equ:prop_p1},
\ref{equ:prop_p2}): 
the non-zero matrix elements among sites $0,1,2$ are
\begin{equation}
G_{01} \simeq \ee^{E_\hf - F}, \quad
G_{21} \simeq \ee^{E_{\frac32} - F},\quad G_{00}=G_{22}=1.
\label{equ:rg_int_G}
\end{equation}
where the approximate equalities simply indicate that these
are propagators under the effective dynamics.
\newcommand{\barr}{}

For the next stage of the effective dynamics, the largest 
hopping rates are $r_0'$ and $\ell_2'$.
Because $E_2>E_0$, the former will be
larger than the latter, and the remaining rates $\ell_0$ and
$r_2$ are much smaller by our assumption (\ref{equ:eg_rates}).
Correspondingly, when we calculate rates $\rho$ and $\lambda$
for the next stage of the effective dynamics, the
largest one will be $\rho_0'=\hf(r_0'+ \ell_0+\ell_2')+\hf\sqrt{
r_0'^2 + 2r_0(\ell_2'+\ell_0)+(\ell_2'-\ell_0)^2}$.
From the assumptions (\ref{equ:eg_rates}),
we have $\ell_0'\ll \ell_2$, and the rules of
the effective dynamics state that we merge sites $0$
and $2$.  Physically, the site energies $E_0$ and $E_2$
are similar so these are combined into an effective
trap which also contains the intervening site 1.
In accordance with (\ref{equ:p_slow}),
the slow degrees of freedom are now simply
$\{\dots,p_{012}(t),\dots\}$ where
$p_{012}(t)=p_0(t)+p_1(t)+p_2(t)$
is the occupancy of an effective trap containing
sites $0$, $1$ and $2$.  The corresponding
energy landscape is shown in Fig.~\ref{fig:rg_new}c.
The equation of motion
for $p_{012}(t)$ at this stage is 
\begin{equation}
\partial_t p_{012}(t) = r_{-1} p_{-1}(t) + \ell_3 p_3(t) -
(r_{012} + \ell_{012}) p_{012}(t) 
\end{equation}
where the new rates appearing at this stage are
$r_{012}=\ee^{-F_{012}-E_{\frac52}}$ for hops to the right
from the effective trap, and
$\ell_{012}=\ee^{-F_{012}-E_{-\frac12}}$ for hops to the
left.  Here,
$F_{012}$ is the trap free energy, 
$\ee^{F_{012}}=\ee^{E_0}+\ee^{E_1}+\ee^{E_2}$,
in accordance with (\ref{equ:Fa}).  
The approximate propagator on time scales
$1/\rho_0'$ can be constructed from (\ref{equ:prop_p1},
\ref{equ:prop_p2}), giving
\begin{equation}
G_{ji}=\ee^{E_j-F_{012}}, \quad 0\leq i,j\leq 2. 
\label{equ:rg_fin_G}
\end{equation}
which is independent of the initial site $i$ as long
as it is within the trap,
consistent with the idea of local equilibration.

\subsection{Comparison with other schemes}
\label{sec:rg_other}

\begin{figure*}
\hfill \includegraphics[width=0.8\textwidth]{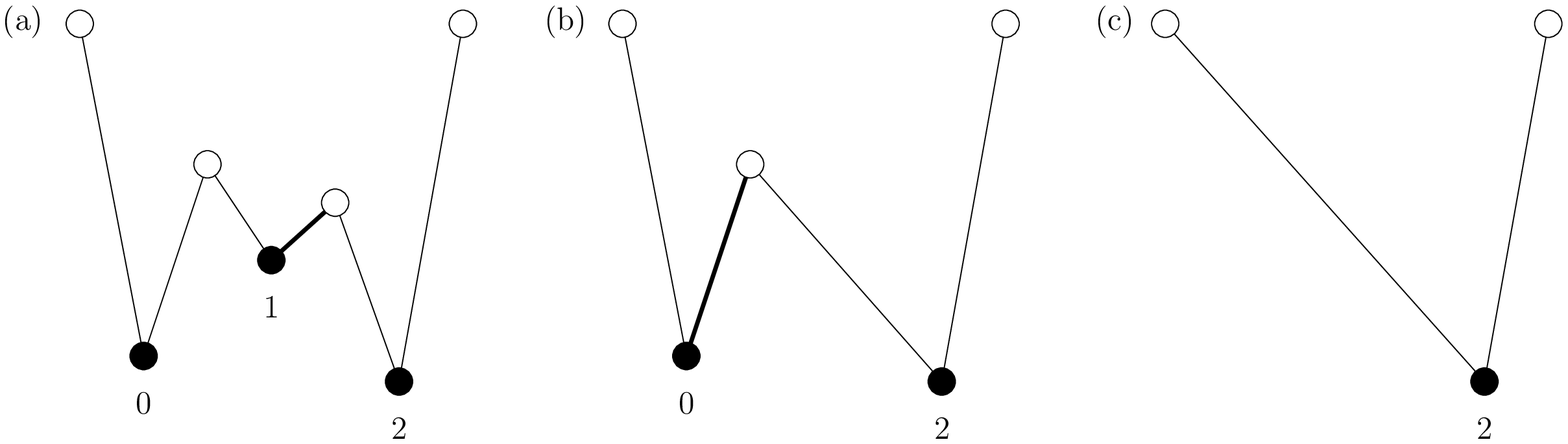}
\caption{Illustration of the RG scheme of DMF, discussed in the main
text.  (a) Energy landscape associated with a master operator
satisfying~(\ref{equ:eg_rates}).
The fastest rate $r_1$ is associated with the energy
difference $E_1+E_{\frac32}$: the link associated with
this rate is emphasised in bold.  (b) Energy landscape
associated with the model after one stage of renormalisation
using the effective
dynamics of DMF.  The fast motion on these time scales
involves a particle initially on site $1$
relaxing onto site $2$, with probability unity.
The link associated with the largest rate 
of the resulting master operator is again emphasised.
(c) Energy landscape after two stages
of renormalisation following DMF.
Particles initially on sites $0$, $1$ or $2$ 
will relax onto site $2$.
}
\label{fig:rg_dmf}
\end{figure*}

It is useful to compare this scheme with the effective
dynamics of DMF~\cite{DMF}.  In that scheme, the 
slow degrees of freedom $p_\alpha(t)$ are simply a subset of
the original site occupancies $p_i(t)$, so each effective
trap contains exactly one site: $a_\alpha=b_\alpha$
for all $\alpha$.  Further, instead
of calculating transition state free energies as
in (\ref{equ:Fahf}), one takes simply 
$\ee^{F_{\alpha-\hf}}=\max_{b_{\alpha-1}\leq j<a_\alpha} \ee^{E_{j+\hf}}$,
assuming that the transition state free energy of each barrier region
is dominated by the largest transition state energy within that region.
Finally, one also takes simply $\lambda_\alpha=\ell_\alpha$
and $\rho_\alpha=r_\alpha$ when deciding which traps to remove as
$\Gamma$ is reduced.
The progress of the scheme is illustrated in 
Fig.~\ref{fig:rg_dmf} for the same example landscape considered
in Fig.~\ref{fig:rg_new}.  
The first step of the renormalisation
scheme takes place when $\Gamma=r_1$ and the slow
degrees of freedom after this step are $\{\dots,p_0(t),p_2(t),\dots\}$
as in our scheme.  In the DMF scheme, the equation
of motion for these degrees of freedom is of the same
form as (\ref{equ:rg_int_eg}),
but the new
rates are $r_0'=\ee^{-E_{\hf}-E_0}$ and $\ell_2'=\ee^{-E_\hf-E_2}$
[note that $E_\hf>E_{\frac32}$, from (\ref{equ:eg_rates}), which
sets the transition state free energy]. 
Further, in the DMF scheme, all elements of the approximate propagator 
are zero or unity: for times of order $1/r_1$, the non-zero
elements of the approximate $G_{ij}$ for $0\leq j\leq 2$ are 
\begin{equation}
G_{00}\simeq G_{21}\simeq G_{22} \simeq 1.
\label{equ:rg_int_G_dmf}
\end{equation}
which can be compared with (\ref{equ:rg_int_G}).

Then, the second
stage of the scheme takes place at $\Gamma=r_0'$,
after which the slow degrees of are $\{\dots,p_2(t),\dots\}$,
and the new rates for motion from the trap containing
site $2$ are $\ell_2''=\ee^{-E_{-\frac12}-E_2}$ and 
$r_2''=\ee^{-E_{\frac52}-E_2}$.  For times of order $1/r_0'$,
the non-zero elements of the approximate propagator between
sites $0,1,2$ are 
\begin{equation}
G_{20}=G_{21}=G_{22}=1
\label{equ:rg_fin_G_dmf}
\end{equation}
which can be compared with (\ref{equ:rg_fin_G}).

In general then, the DMF scheme gives different
results to our scheme, although broad features are similar.  
In the limit where all rates are well-separated:
$r_1\gg \ell_1\gg l_2\gg r_0 \gg \ell_0,r_2$, 
it may be verified that the two schemes coincide.
That is the condition in which the DMF scheme is exact, and
we conclude that our scheme is also exact in that limit.
In fact, our scheme is also exact in the less-restricted
limit of well-separated
rates given in (\ref{equ:eg_rates}), while the DMF method is inaccurate
in that case.  The clearest differences between the schemes
occur in the propagators.
For example, compring (\ref{equ:rg_int_G}) with (\ref{equ:rg_int_G_dmf}),
the DMF scheme ignores the possibility that 
a particle originally on site $1$ may relax onto site $0$ at
this stage, while such transitions
happen with probability approaching $\hf$ if $r_1\approx \ell_1$.
Additionally, under the same condition $r_1\approx \ell_1$,
the transition
rates $r_0'$ and $\ell_2'$ differ between the schemes by a factor
close to 2.  Compared
to DMF, these rates are smaller in our new scheme, reflecting
the possibility that a particle that hops from site $0$ to site
$1$ may return to site $0$ before visiting site $2$.  As discussed
in \cite{JS-dual}, factors such as these must be included in
effective dynamics schemes in order to obtain the correct
scaling behaviour for models where many of the $E_{i+\hf}$
are approximately equal.

We also compare the scheme given here with that of 
MG~\cite{Garel-Monthus}.  That scheme states the slow
co-ordinates at each stage and their equations of motion, although
the explicit propagator $G_{mn}$ among the original sites of
the model is not given.  In fact, for one-dimensional
hopping models, the `full' MG scheme reduces to a simplified version
of our scheme, in which effective traps are always simply
removed from the list of slow degrees of freedom, but traps
are never merged.  In the language of Sec.~\ref{sec:rg},
one always assumes that the landscape has a `pure trap'
character.  In choosing which traps
to remove, one takes the parameters 
$\lambda_\alpha=\rho_\alpha=r_\alpha+\ell_\alpha$, consistent
with that assumption.
One may verify that this procedure is
exact and coincides with our scheme
in the limit where all site energies $E_i$ are
well-separated from each other: it is therefore
appropriate for systems such as pure trap models.
However, the scheme does not preserve the duality
of Sec.~\ref{sec:dual}: in particular, while it is
appropriate for 
pure trap models, it fails for their duals, which
are pure barrier models.

\subsection{Operator representation of effective dynamics}
\label{sec:separation}

The effective dynamics scheme can be interpreted as
a projection of the master operator $W$ onto its slow
degrees of freedom.  
Briefly,
any master operator can be diagonalised as $W=-\sum_\lambda|\lambda_R\rangle
\lambda\langle\lambda_L|$ and its propagator written
as $\ee^{Wt}=\sum_\lambda|\lambda_R\rangle
\ee^{-\lambda t}\langle\lambda_L|$.  Time scales are (globally)
well-separated if
there is a time $t$ such that all of the exponential factors
are either negligibly small or close to unity.  For
times $t$ with that property, 
the time evolution operator 
is well-approximated by a projection operator
\begin{equation}
\ee^{Wt}\approx\mathcal{P}_\mathrm{ex}(t^{-1})\equiv
\sum_{\lambda<t^{-1}} |\lambda_R\rangle \langle \lambda_L|,
\label{equ:Pex}
\end{equation}
which can be rearranged into the form
$\ee^{Wt}\approx\sum_{\alpha} |p_\alpha\rangle \langle q_\alpha|$
where the states $|p_\alpha\rangle$ and $\langle q_\alpha|$
have non-negative elements~\cite{timescale-sep}.  
For the specific
case of hopping models, the states $|p_\alpha\rangle$ indicate
the effective traps into which the system relaxes
on the time scale $t$, while the vectors $\langle q_\alpha|$ indicate
the probabilities of relaxing into these states. 

For the effective dynamics, the key point is that motion
on time scales longer than $t=\Gamma^{-1}$ can be well-described
by a renormalised master operator
$W_\mathrm{R,ex}(\Gamma) = \mathcal{P}_\mathrm{ex}(\Gamma) W 
\mathcal{P}_\mathrm{ex}(\Gamma)$, since 
eigenmodes with $\lambda t\gg 1$ that are irrelevant at time
$t$ are also irrelevant for all longer times [more precisely,
$\ee^{Wt}=\ee^{W_\mathrm{R,ex}(\Gamma)t}+O(\ee^{-\Gamma t})$ and the
error becomes small for $t\gg\Gamma^{-1}$].
In the effective dynamics a set 
of effective trap and barrier regions corresponds
to an operator $\mathcal{P}(\Gamma)$ that
approximates $\mathcal{P}_\mathrm{ex}(\Gamma)$.  We therefore
define the renormalised master operator
\begin{equation}
W_\mathrm{R}(\Gamma) = \mathcal{P}(\Gamma) W \mathcal{P}(\Gamma)
\end{equation}
and we note that a corresponding approximation for the propagator
is
\begin{equation}
G_{mn}(\Gamma^{-1}) \simeq \langle m|\ee^{W_R/\Gamma}|n\rangle
 \simeq \langle m| \mathcal{P}(\Gamma) |n\rangle
\end{equation}

For consistency with (\ref{equ:master_slow}) above, the renormalised
master operator should take the form
\begin{equation}
W_\mathrm{R}(\Gamma) 
= \sum_{\alpha} 
( |P_{\alpha+1}\rangle - |P_\alpha\rangle ) 
( r_\alpha \langle Q_\alpha |  - \ell_{\alpha+1} \langle
Q_{\alpha+1} |)
\label{equ:wr_def}
\end{equation}
where $|P_\alpha\rangle$ and $\langle Q_\alpha|$ are 
vectors associated with trap $\alpha$.  Thus,
the effective dynamics represent a renormalisation scheme
in the sense that the operator $W_\mathrm{R}$ maintains
the same form as the original master operator~(\ref{equ:Wt})
as the cutoff $\Gamma$ is reduced.
To obtain such a form for $W_\mathrm{R}$, we take
\begin{equation}
\mathcal{P}(\Gamma) = \sum_\alpha |P_\alpha\rangle \langle Q_\alpha|
\end{equation}
where the $|P_\alpha\rangle$ and $\langle Q_\alpha|$ are
approximations to slow eigenvectors of $W$.  
Our scheme as given by Equs.~(\ref{equ:p_slow}-\ref{equ:rla}) 
corresponds to the choice
\begin{equation}
\langle i | P_\alpha\rangle = \ee^{E_i-F_\alpha}
\label{equ:Prg}
\end{equation}
and
\begin{equation}
\langle Q_\alpha | i \rangle = \left\{ \begin{array}{ll}
 v_i^{(\alpha-\hf)}, & b_{\alpha-1}<i<a_\alpha \\
 1, & a_\alpha \leq i \leq b_\alpha \\
 1-v_i^{(\alpha+\hf)},\quad & b_\alpha < i< a_{\alpha+1}
\end{array} \right.
\label{equ:Qrg}
\end{equation}
In \ref{app:eff}, we show that the $|P_\alpha\rangle$
and $\langle Q_\alpha|$ constructed in this way are
indeed good approximations to slow eigenvectors of $W$,
under conditions discussed below.

Thus, the effective equation of motion (\ref{equ:master_slow})
corresponds in operator notation to the 
equation $\partial_t |P(t)\rangle = W_R(\Gamma) |P(t)\rangle
= \mathcal{P}(\Gamma) W \mathcal{P}(\Gamma) |P(t)\rangle$
and the approximate propagator of (\ref{equ:prop_p1},\ref{equ:prop_p2}) 
corresponds
to $G_{mn}(\Gamma^{-1})\simeq \langle m|P(\Gamma)|n\rangle$.
The quality of these approximations depends 
on two considerations.  Firstly, the validity of the effective
equation of motion (\ref{equ:master_slow}) depends on
the extent to which the projection operator
${\cal P}(\Gamma)$ approximates ${\cal P}_\mathrm{ex}(\Gamma)$.
Then, the extent to which the propagator $G_{mn}(\Gamma^{-1})$ 
coincides with $\langle m|P(\Gamma)|n\rangle$ depends in addition
on a separation of time scales, as can be seen from the discussion 
of (\ref{equ:Pex}).

We observe that if ${\cal P}(\Gamma)\approx{\cal P}_\mathrm{ex}(\Gamma)$
at all stages in the scheme then the errors associated with the effective
dynamics remain small as $\Gamma$ is reduced, while large errors arise if
${\cal P}(\Gamma)$ becomes different from ${\cal P}_\mathrm{ex}(\Gamma)$.
The conditions under which the operator $\mathcal{P}(\Gamma)$ 
represents a good approximation to $\mathcal{P}_\mathrm{ex}(\Gamma)$
are discussed
in~\ref{app:eff}.  Some numerical tests are also given in Sec.~\ref{sec:numerics}.
We summarise here the analytic results of~\ref{app:eff}:
On long time scales (small $\Gamma$) we identify the `fastest
relevant rates' $\ell_\alpha$ and $r_\alpha$ 
which are comparable to $\Gamma$.  For a consistent
renormalisation flow, we require that
these fast relevant rates are typically 
much larger than all other relevant rates in their neighbourhood,
except that (i) fast relevant rates $r_\alpha$ may be comparable 
either to $\ell_\alpha$ or to $\ell_{\alpha+1}$, and
(ii) fast relevant rates $\ell_\alpha$ may be comparable either to
$r_\alpha$ or to $r_{\alpha-1}$. 
(The condition of globally well-separated
time scales described above is not required: it is sufficient
that eigenvalues for motion in the same spatial neighbourhood
should be well-separated.)

\subsection{Renormalisation and duality}
\label{sec:renorm-dual}

To conclude this section, we discuss duality relations for 
the effective dynamics, restoring superscripts to distinguish between
$\Wint$ and $\Whf$.
If we renormalise an operator
$\Wint= S \Sbar$ according to our scheme, we arrive at a renormalised
model that can be written in the form $\Wint_\mathrm{R}=S_\mathrm{R}
\Sbar_\mathrm{R}$. The operators 
$S_\mathrm{R}$ and $\Sbar_{\mathrm{R}}$ have the same form as $S$ and
$\Sbar$, except that sites $i$ are replaced by effective traps
$\alpha$, transition states by effective barriers, and energies by free energies.
An important property of our RG procedure is that if we apply it to
the dual master operator $\Whf=\Sbar S$, we find that this 
renormalises precisely to 
$\Whf_\mathrm{R} = \Sbar_\mathrm{R} S_\mathrm{R}$.
Thus, in addition to the basic
requirement that $W_\mathrm{R}(\Gamma)$ takes the same form as
$W$, our scheme also obeys the general duality relation under
landscape inversion. This is of course desirable, as renormalisation
schemes should respect all symmetries of the models of interest.  

In the illustrations of Figs.~\ref{fig:rg_new} and~\ref{fig:rg_dmf},
the duality property follows for both the new scheme and that of DMF,
because acting on these illustrations with
the inversion operation of Fig.~\ref{fig:invert_all} leads to
the same renormalisation flows that would be obtained by starting
with the dual of the original model.  

Mathematically, the duality can be shown as follows. At each stage of
the RG flow we have effective trap regions associated with
$\Wint_\mathrm{R}$, with intervening barrier regions.  Assigning
integer indices $\alpha$ to the traps, the barriers can
be associated with indices $\alpha+\hf$.
In $\Whf_\mathrm{R}$ the regions with integer indices $\alpha$ become
effective barriers, while those with indices $\alpha+\hf$ become
effective traps.  The associated free energies $F_\alpha$ and
$F_{\alpha+\hf}$ are the same in both cases. In the dual model
$\Whf_\mathrm{R}$, the rates for hopping to right and left from
trap $\alpha-\hf$ are
$r_{\alpha-\hf}=\ell_\alpha$ and $\ell_{\alpha-\hf}=r_{\alpha-1}$,
consistent with (\ref{equ:LRlr}).
One also easily checks that the rates $\lambda_{\alpha+\hf}$ 
and $\rho_{\alpha+\hf}$
that are obtained on renormalising $\Whf$ are the same
as those obtained on renormalising $\Wint$, 
according to~$\rho_{\alpha-\hf}=\lambda_\alpha$ and
$\lambda_{\alpha+\hf}=\rho_\alpha$.  Thus, supposing that
we combine traps $\alpha$ and $\alpha+1$ in an RG step
on the model $\Wint$, we also remove barrier $\alpha+\hf$.
In the dual model $\Whf$, we remove the trap with index $\alpha+\hf$,
which corresponds to combining the barrier regions $\alpha$ and $\alpha+1$.
Finally, it can be verified that 
be verified that the rules for merging and removing
traps do preserve the duality between $\Wint_\mathrm{R}$
and $\Whf_\mathrm{R}$.

A brief comment is in order on the construction of $S_\mathrm{R}$ and
$\Sbar_\mathrm{R}$. From $\Wint_\mathrm{R}=\mathcal{P}^{(1)} \Wint
\mathcal{P}^{(1)}$ and $\Wint=S\Sbar$ one might naively identify
$S_\mathrm{R}=\mathcal{P}^{(1)}S$,
$\Sbar_\mathrm{R}=\Sbar\mathcal{P}^{(1)}$; but this choice does not
satisfy the duality requirement that $\Whf_\mathrm{R} =
\Sbar_\mathrm{R} S_\mathrm{R}$. A little thought shows that one
requires instead $S_\mathrm{R}=\mathcal{P}^{(1)}S\mathcal{P}^{(1/2)}$
and $\Sbar_\mathrm{R}=\mathcal{P}^{(1/2)}\Sbar\mathcal{P}^{(1)}$. Here
$\mathcal{P}^{(1/2)}=\sum_\alpha |P_{\alpha+\hf}\rangle\langle
Q_{\alpha+\hf}|$ is the projector onto the effective traps of the dual
model, with $|P_{\alpha+\hf}\rangle$ and $\langle Q_{\alpha+\hf}|$
constructed in the obvious manner using the duals of (\ref{equ:Prg})
and (\ref{equ:Qrg}). The operators $S_\mathrm{R}$ and
$\Sbar_\mathrm{R}$ defined in this way are indeed the effective trap
and barrier analogues of (\ref{eq:S1st}--\ref{eq:Sbar2nd}), and the
desired dual expressions $\Wint_\mathrm{R}=S_\mathrm{R}
\Sbar_\mathrm{R}$ and $\Whf_\mathrm{R} =
\Sbar_\mathrm{R} S_\mathrm{R}$ therefore hold. We note finally that
not only the master operators but also the propagators produced by our
RG scheme obey the required duality: the propagators are the
matrix elements of the projectors $\mathcal{P}^{(1)}$ and
$\mathcal{P}^{(1/2)}$, and one verifies by direct calculation that
$\langle m|\mathcal{P}^{(1)}(|n\rangle - |n+1\rangle)=\langle
n+\hf|\mathcal{P}^{(1/2)}(|m+\hf\rangle - |m-\hf\rangle)$ in
accordance with the general duality relation (\ref{equ:dual_prop}).

\section{Specific disorder distributions}
\label{sec:numerics}

\subsection{Mixed trap-barrier models}
\label{sec:mixed}

The renormalisation scheme that we have discussed can be implemented 
computationally without undue difficulty.  It allows rapid
estimation of propagators in these hopping models, both
for fixed disorder and for disorder-averaged properties.  We
first consider a model obtained by mixing the pure trap and barrier
models defined above.
In the pure trap model, the transition state energies are
$E_{i+\hf}=0$, while site energies are chosen from an 
exponential distribution with a mean of $\mu$; that is,
 $P(E_i) = (1/\mu) \ee^{-E_i/\mu}$ with
$E_i>0$.  In terms of 
rates, this implies $\ell_i=r_i=w_i$ with $P(w_i)=(1/\mu) w_i^{(1/\mu)-1}$
for $0<w_i<1$.
Similarly, pure barrier models have $E_i=0$ for all sites,
and transition state energies are
exponentially distributed: that is,
$L_i=R_i=w_i$ with the same distribution $P(w_i)$.

We mix these models by taking both site and transition state energies
to be exponentially distributed with means $\mut$ and 
$\mub$ respectively.  
The dynamical scaling of these models therefore depends on the parameters
$\mut$ and $\mub$.  For the pure barrier model ($\mut\to0$),
sites have half-integer indices and 
moving a distance $r$ typically requires the crossing of a barrier $i$ 
whose hopping rate
is $w_i=\ee^{E_{i+\hf}}\sim r^{-\mub}$.  The rate for actually crossing
this barrier is suppressed because the particle is delocalised in
an effective trap whose width is of order $r$.  
In the language of the effective dynamics, the landscape
consists of wide effective traps separated by isolated
transition states, and  each site $i$ within
the trap contributes contributes $\ee^{E_{i}}=1$ to 
$\ee^{F_{\alpha}}$.
The result is that the time
taken to move a distance $r$ is $\tau(r)\sim \ee^{F_{\alpha}+E_{i+\hf}}\sim
r/w_i \sim r^{1+\mu}$.
In the pure trap model, the typical
time for escaping from sites with large $E_i$
is $1/w_i \sim \ee^{E_i}$, but the barrier regions on this
time scale are typically of width $r$ and their free energies
therefore also scale as $\ee^{F_{\alpha+\hf}}\sim r$, reflecting the
probability 
of reabsorption in the original trap before arriving at a new
one~\cite{JS-dual}.  
Thus, the typical relevant time scale is again 
$\tau(r) \sim r^{1+\mu}$. 

We define the dynamical exponent $z$ through
the relation $\tau\sim r^z$, and identify
\begin{equation}
z = 1+\mub, \qquad \mub>1,\, \mut\to0
\end{equation}
with an analogous relation if $\mut>1$ and $\mub\to0$.

\begin{figure}
\begin{center}
\includegraphics[width=0.48\textwidth]{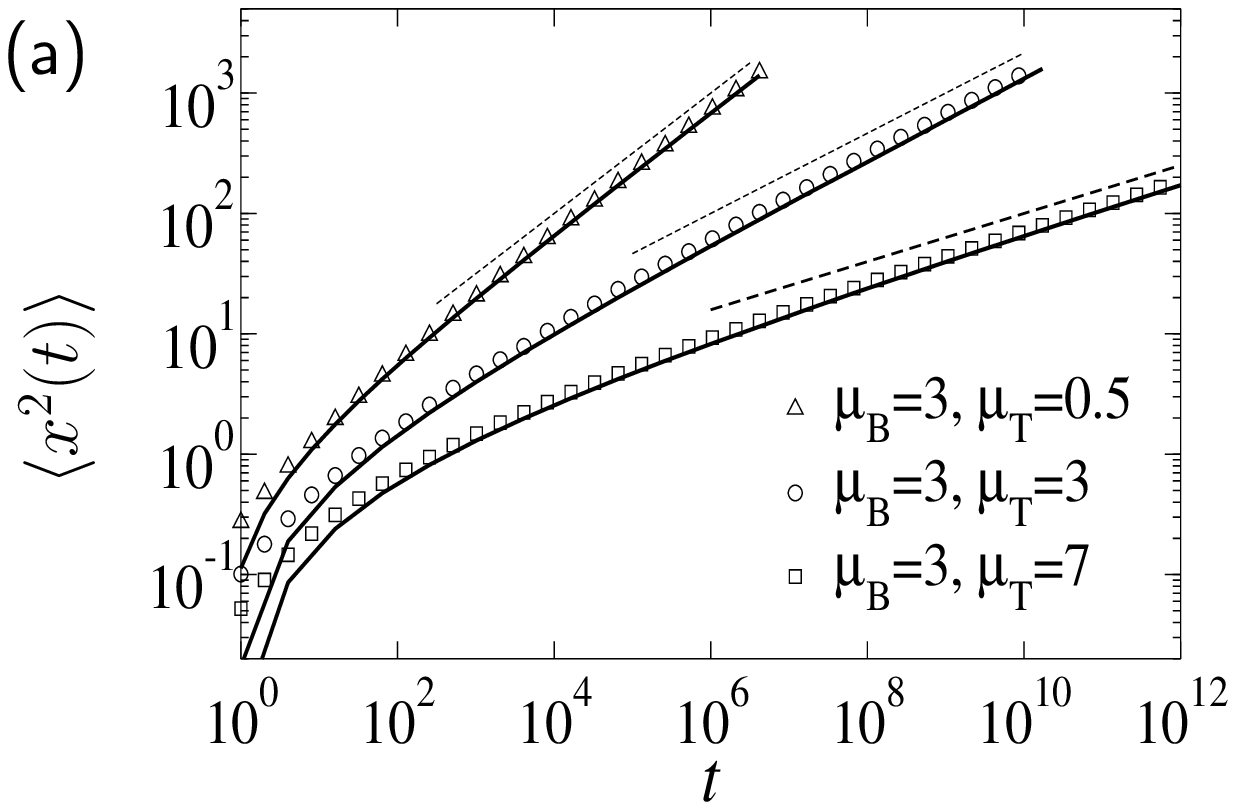}\par
\includegraphics[width=0.48\textwidth]{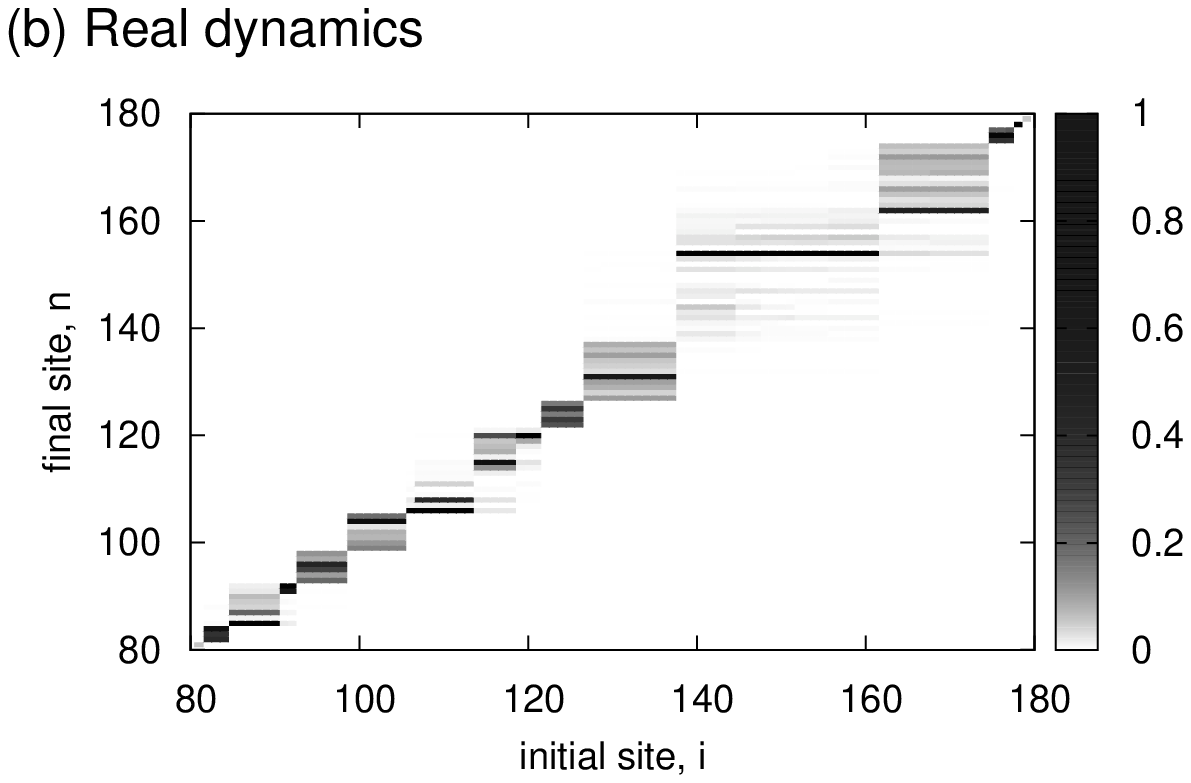}
\includegraphics[width=0.48\textwidth]{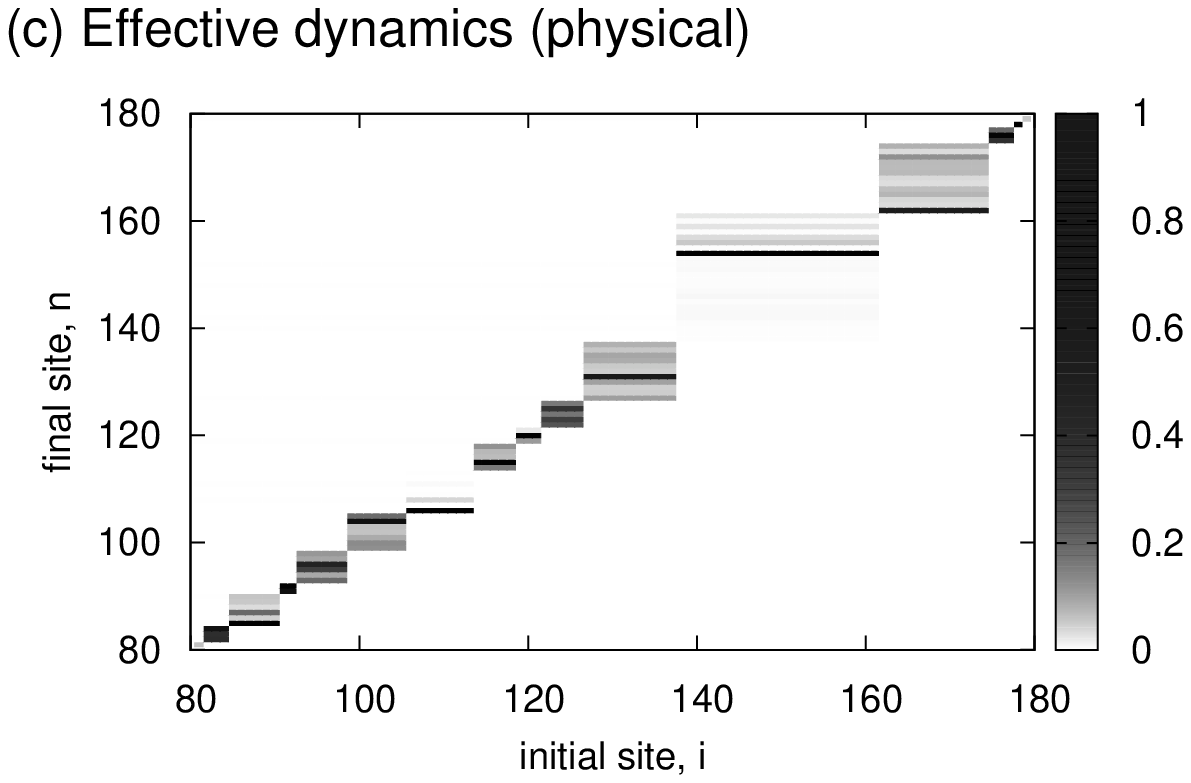}
\includegraphics[width=0.48\textwidth]{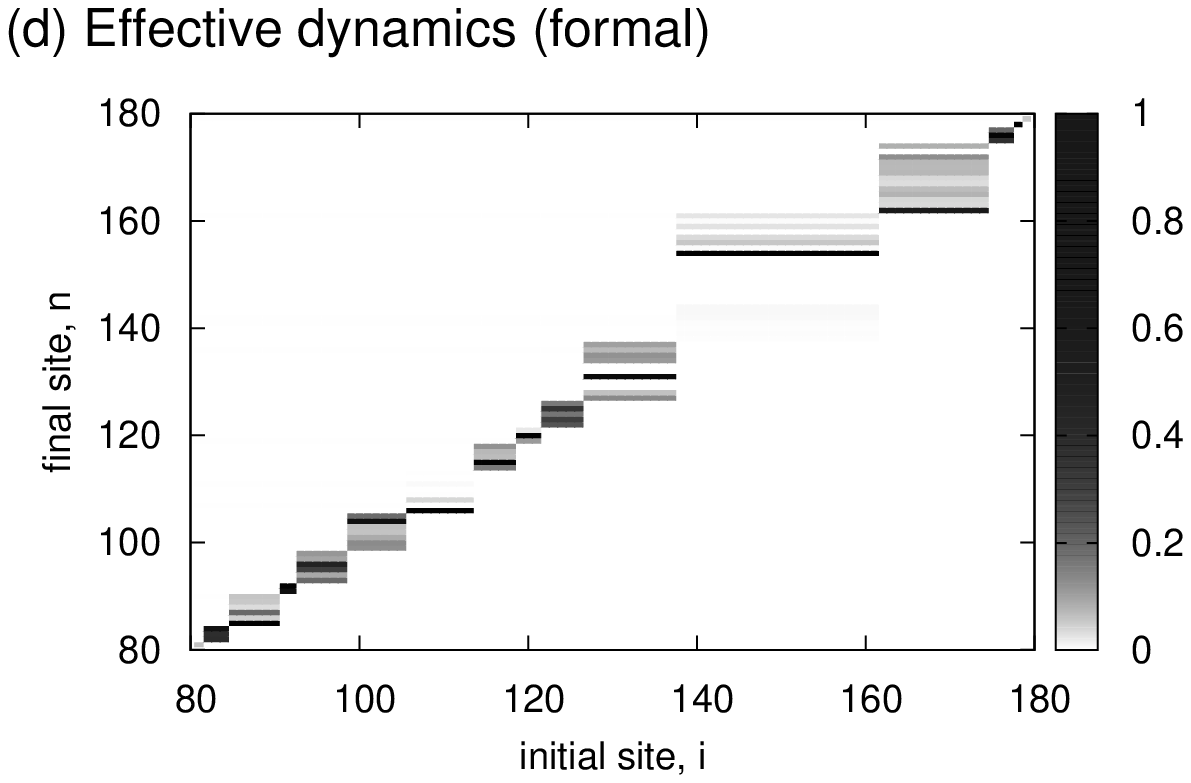}
\includegraphics[width=0.48\textwidth]{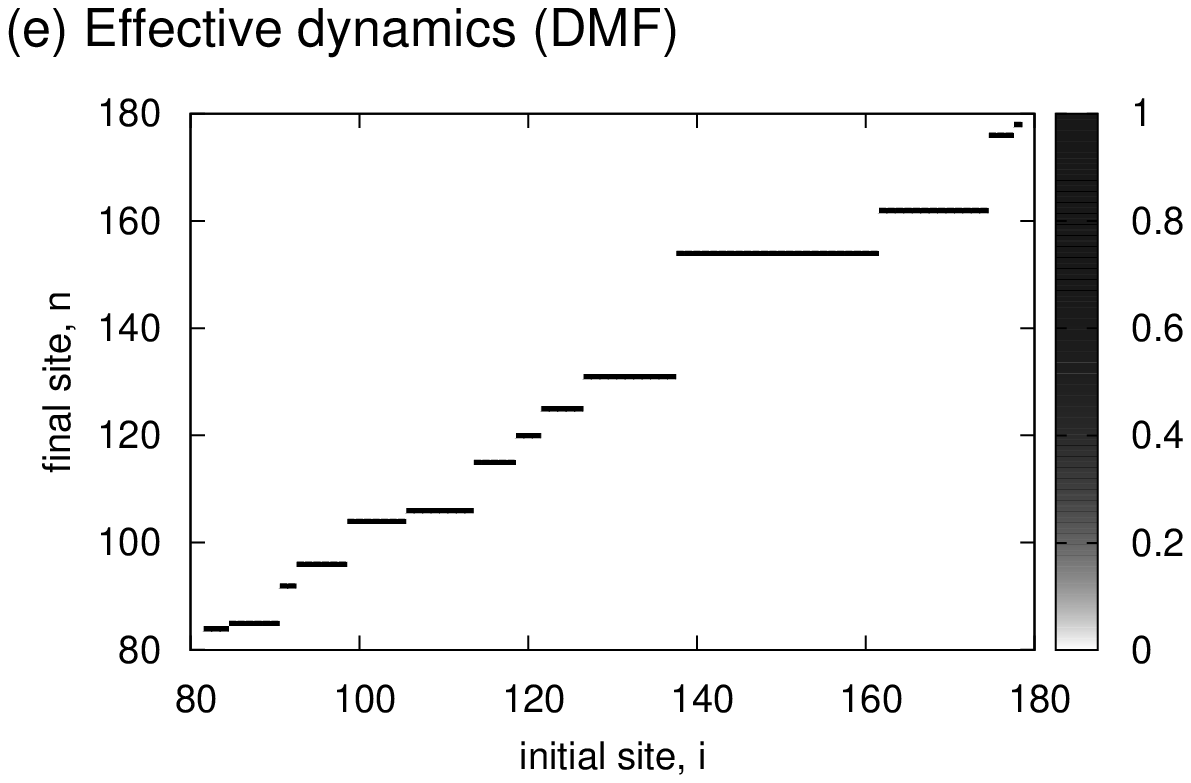}
\end{center}
\caption{(a) Mean square displacement, showing real dynamics
(symbols), effective dynamics (solid lines) and power law
predictions of (\ref{equ:z-mixed}) (dashed lines).  The
DMF scheme (not shown) also gives the correct scaling
as long as $\mub,\mut>1$ but gives the wrong exponent
for the case where $\mut=0.5$.  (b,c)~Propagators for
real dynamics~(b) and effective dynamics~(c) for
the case $\mub=10$ and $\mut=1$.
The time is $t=2^{29}$: the propagator for the
real dynamics is obtained by simulating its dual model
and using (\ref{equ:dual_prop}), since models with
large $\mut$ can be simulated much more efficiently
than those with large $\mub$.
Since $\mub$ is large,
nearby rates are well-separated and the effective
dynamics gives a good approximation to the propagator even
for fixed disorder. However, there are deviations in some
neighbourhoods which are
associated with the presence of eigenvalues of the
same order as $1/t$, as discussed in the text.
(d,e)~Propagators for the formal effective
dynamics scheme (see \ref{app:eff}) and the DMF scheme
(see Sec.~\ref{sec:example_rg}).  The formal
scheme differs slightly from the physical scheme:
for example, the formal scheme gives $G_{ij}=0$ for $i=131,132$,
independent of $j$, while
the physical scheme gives small finite values that
are more consistent with the real dynamics.  In the DMF
scheme, all effective traps consist of single sites,
so the scheme does not capture the broad `effective traps'
that are visible in the real dynamics. 
}
\label{fig:mix-num}
\end{figure}

In the mixed model, the crucial case distinctions are then
whether the $\mub$ and $\mut$ are larger or smaller than unity.  For
example, if $\mut<1$, the average value of $\ee^{E_i}$ is finite,
and the free energies of relevant trap regions
scale as $\ee^{F_i}\sim r$.  On the other hand,
if $\mut>1$, the site-averaged $\ee^{E_i}$ is no longer finite,
and the sum in (\ref{equ:Fa}) is dominated by the largest site within the 
effective trap.  In this case $\ee^{F_i}\sim r^{\mut}$.
A similar argument applies for $\mub$.  Combining
these results, we arrive at the dynamic exponent for the
mixed model
\begin{equation}
z  = \mathrm{max}(1,\mub) + \mathrm{max}(1,\mut)
\label{equ:z-mixed}
\end{equation}
which reduces to the pure trap  case if $\mub=0$
and the pure barrier case if $\mut=0$.  Fig.~\ref{fig:mix-num} shows
numerical results that are consistent with~(\ref{equ:z-mixed}).
Thus, the effective dynamics provide a natural framework
in which to derive this kind of scaling result, although the
above predictions for the dynamical exponent could presumably be
obtained by other means. 
We also note that the renormalisation arguments
of DMF give $z=\mub + \mut$ which
is the correct result when both $\mub$ and $\mut$
are greater than unity.  This is consistent
with our assertion above that if $\mub,\mut>1$ the free energies
of effective traps and barriers are typically dominated by single
sites, and this is precisely the limit in which the scheme of
DMF is valid without approximation.  To be precise,
it follows from the discussion
of~\ref{app:eff} that our effective dynamics scheme is exact
in the limit where either $\mub$ or $\mut$ is very large,
while the DMF scheme is exact in the limit where both $\mub$
and $\mut$ are very large.  

In Fig.~\ref{fig:mix-num}(b,c),
we show example propagators obtained using real and effective
dynamics for the case $\mub=10$, $\mut=1$.
For these parameters,
time scales in a given neighbourhood are sufficiently
well-separated that the effective dynamics gives a good approximation
to the propagator.  In this case, 
it appears that the largest deviations between real and effective
dynamics come from neighbourhoods in which there is an eigenvalue
of $W$ of the order of $1/t$.  
In the discussion of
Sec.~\ref{sec:separation}, we noted that such deviations are
expected even when our scheme gives $P(\Gamma)$ exactly
equal to $P_\mathrm{ex}(\Gamma)$, and that the specific
deviations seen at any
time $t$ should decay as time increases, so that the exact
and approximate propagators remain close.  Our results are
consistent with this expectation.
For example, comparing Figs.~\ref{fig:mix-num}b and~\ref{fig:mix-num}c,
the effective dynamics indicates that
sites in the vicinity of site 110 are separated into
three traps, containing sites 106-113, 114-118, and 119-121:
the effective barrier regions are simply single transition states.
However, the real dynamics reveals that initial sites $114-118$ 
propagate both within that effective trap, and into
the adjacent traps.  On time scales much shorter than $t$, one would
expect localisation within this trap; on longer time
scales the trap will either merge with an adjacent
trap, or become incorporated into an effective barrier region.  We
conclude that we have measured the propagator during this crossover,
the details of which are not captured by the effective dynamics.

In Fig.~\ref{fig:mix-num}(d,e), we compare the physical
effective dynamics of~\ref{sec:rg} with the formal
scheme of~\ref{app:eff} and the scheme of DMF.  As discussed
in Sec.~\ref{sec:rg_other}, the DMF scheme assumes that all
effective traps consist of only a single site, so for any
initial site $i$, there is a single final site $j$ such that $G_{ji}=1$,
with $G_{ji}=0$ for all other final sites.  It can be seen that
the DMF scheme does identify final sites $j$ with large $G_{ji}$, but it
underestimates $G_{ji}$ for other $j$.  As discussed in~\ref{app:eff},
our physical effective dynamics scheme means that for a given $i$,
$G_{ji}$ is finite for $j$ within contiguous regions of the chain; 
on the other hand, the formal scheme leads to $G_{ji}$ that is finite 
on a restricted set of sites within such regions.  As discussed
in the caption to Fig.~\ref{fig:mix-num}, the formal
scheme therefore underestimates $G_{ji}$ for some sites $j$. 
Since $\mub=10$ is quite large, we expect both of our effective
dynamics schemes to mimic the real dynamics quite accurately,
consistent with the data.  Deviations between the two schemes
and the real dynamics would increase as $\mub$ is reduced (data not
shown).  Similarly, 
the predictions of the DMF would mimc the real dynamics more
closely if $\mut$ were increased since that scheme requires
both $\mut$ and $\mub$ to be large.

\newcommand{\rr}{k}

\begin{figure}
\begin{center} \includegraphics[width=8cm]{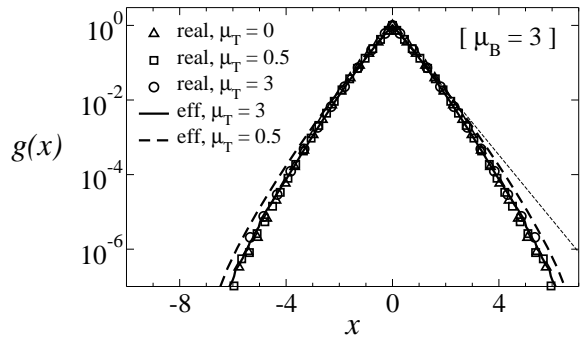}
 \end{center}
\caption{
We show estimates of the function
 $g(x)$ obtained by evaluating $\Gbar_{\rr}(t)$ for
times $t$ within
the scaling regime. Results are displayed
for both real and effective dynamics, with $\mub=3$.
Symbols show results for real dynamics with $\mut=0$ (triangles),
$\mut=0.5$ (squares) and $\mut=3$ (circles).  The times
are chosen to lie in the scaling regime and vary from $2^{12}$
to $2^{24}$, according to the model:
we note that while the functions $g(x)$ are similar in all cases,
the values and scalings of the mean square displacements are different,
according to~(\ref{equ:z-mixed}).
The solid line shows the result for the effective dynamics
and $\mut=\mub=3$, while the heavy dashed
line shows effective dynamics for $\mub=3$, $\mut=0.5$.
The light dashed line
is a simple exponential distribution. The scheme of DMF (not shown)
gives reasonable agreement for $\mut=\mub=3$, but
this agreement breaks down as $\mut$ (or $\mub$) is reduced.}
\label{fig:fronts}
\end{figure}

 Moving to disorder-averaged properties, numerical results 
 indicate that the long-time behaviour in these systems is
 associated with a scaling form of the diffusion front, as 
 expected.  That is (with $\rr$ the distance between initial and final
site as before),
\begin{equation}
{\Gbar_\rr}(t) \approx 
      \sigma^{-1} g(\rr/\sigma).
\end{equation} 
 where the function $g(x)$ is independent of the time
  $t$ and we use $\sigma=\sigma(t)=1/{\Gbar_0}(t)$ as an estimate
 of the length scale associated with motion on a time scale $t$,
 ensuring that $g(0)=1$. 
 Our numerical results then indicate that
 the shape of the diffusion front $g(x)$
 depends quite strongly on $\max(\mub,\mut)$ and much
  more weakly on $\min(\mub,\mut)$.
 As in Ref.~\cite{JS-dual}, the effective dynamics
 give good agreement with the real dynamics
 when either $\mub$ or $\mut$
 is large, with deviations at smaller $\mu$ that arise because
 time scales associated with hopping rates in the same neighbourhood
 are not well-separated.  We show some illustrative results 
 in Fig.~\ref{fig:fronts}: the fit for the effective dynamics
 with $\mub=\mut=3$ is strikingly good.  However,
 reducing the value of $\mut$ further
 has an effect on the diffusion front for the effective dynamics,
 while no effect is discernable for the real dynamics. This reduces
 the quality of the fit in this case.  
(The results shown are for the physical effective
dynamics scheme. For these disorder-averaged
quantities, we note in passing that the differences
between our `physical' and `formal' 
schemes are of the same order as the differences
bewteen real and effective dynamics,
with the physical scheme being slightly closer to the real
dynamics than the formal one.)

 We also find that numerical implementation of the DMF
 procedure yields reasonable agreement with mean-square
 displacement and the diffusion
 front for the case $\mub=\mut=3$.  However,
 this agreement breaks down as $\mut$ is reduced: 
 for $\mut<1$, the DMF scheme yields the wrong dynamical
 exponent, as discussed above.
 Thus, the
 main advantage of the scheme presented here is that it captures
 the crossover as $\mut$ gets small [see
 Eq.~(\ref{equ:z-mixed})]; in this case, it
also gives more accurate results for the
propagators at fixed disorder (recall Fig.~\ref{fig:mix-num}(c,d,e)).

Finally, we note that when calculating propagators, we 
expect the various schemes (DMF, Mon03, that of \cite{JS-dual},
 and the one
presented here) to be equivalent in the limit of large
$\mu$, at least at the level of the diffusion front.   
However, the approach to that limit is non-trivial
and involves effects that are non-perturbative in $\mu$: the scheme
presented here captures some of these effects, which results
in improved fits to the diffusion front.  
In the next section, we illustrate this in the case of
the pure trap 
model. Of course, because of duality, an essentially identical
discussion can be given for the pure barrier case.

\subsection{Comparison of RG schemes in the pure trap models}

\begin{figure}
\includegraphics[width=0.48\textwidth]{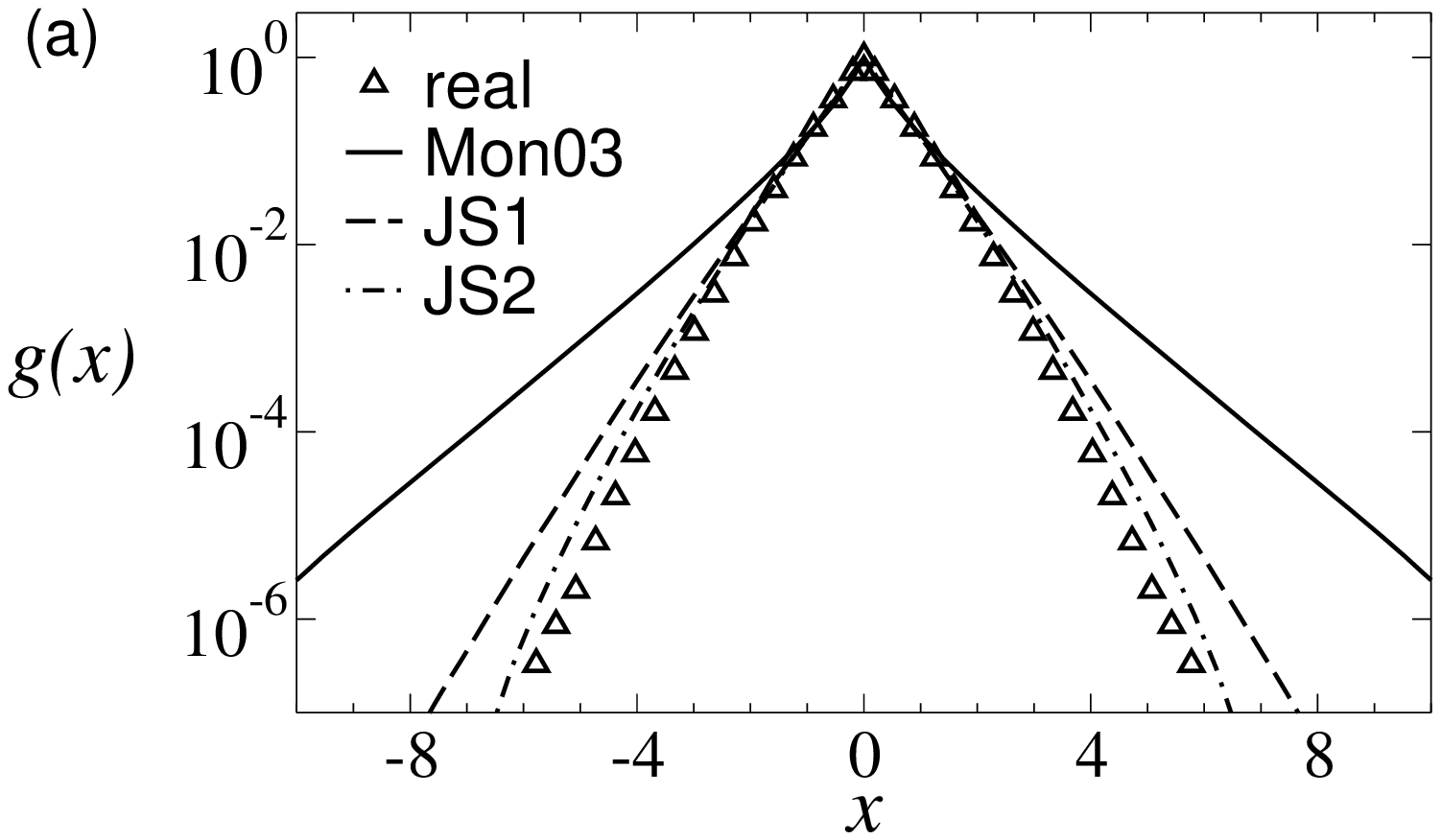}
\includegraphics[width=0.48\textwidth]{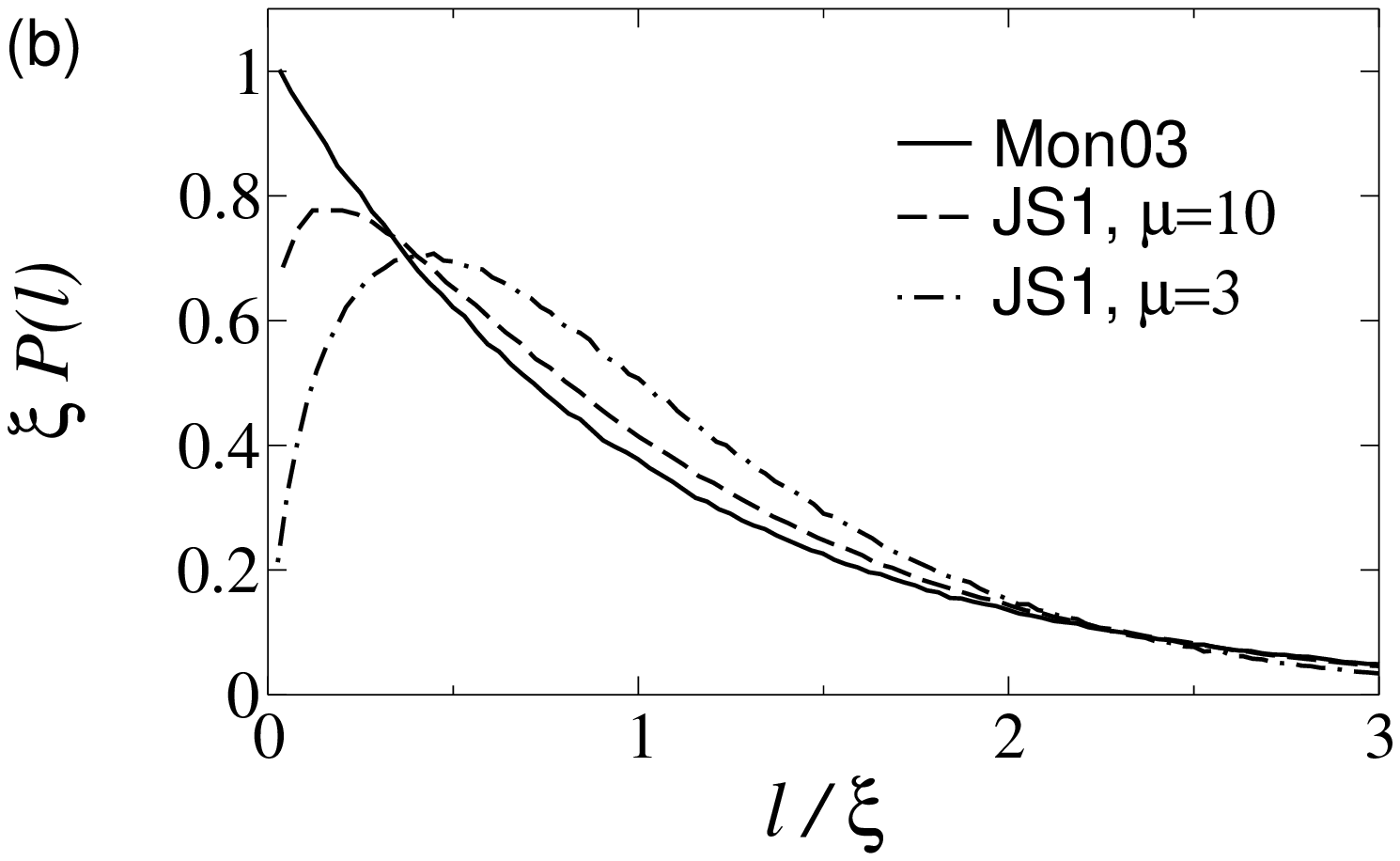}
\caption{(a) Scaling form for the disorder-averaged
diffusion front 
for the pure trap model with $\mut=3$, $\mub=0$. 
We show the results for real dynamics 
and various approximations to this function, obtained
by simulations of effective dynamics schemes. All results are taken
from the scaling regime, with in particular $t=2^{17}$ for the real dynamics.
We show the Mon03 scheme of Ref.~\cite{Mon03}, which yields
(\ref{equ:mon_diff}); the scheme
of \cite{JS-dual} (labelled JS1); 
and the scheme discussed in this article (labelled JS2).
(b) Distribution of barrier widths $P(l)$ in the effective dynamics, scaled by 
its mean $\xi$.  We show results for the effective dynamics scheme
of Mon03 in which $P(l)$ is a simple exponential, and
two distributions obtained with the scheme JS1.}
\label{fig:barr_eff}
\end{figure}

We recall that the pure trap model is the case $\mub\to0$;
we then write $\mut=\mu$.
Applying the DMF method directly to the pure trap model 
results in a dynamical exponent $z=\mu$.  As noted in~\cite{Mon03},
this result is incorrect: the route taken by Monthus was simply
to introduce a factor of the root mean square displacement when
converting the rate $\Gamma$ to a time $t$.  With this change,
the dynamical scaling given by the Mon03 scheme~\cite{Mon03} is correct.  

In Fig.~\ref{fig:barr_eff}, we compare the effective dynamics scheme
set out
in this article with that of \cite{JS-dual} and with Mon03 \cite{Mon03}.
The differences are quite striking, as we now discuss.
In the schemes of~\cite{JS-dual} and~\cite{Mon03}, all
effective trap regions are single sites, and the
RG scheme relates the propagator to the distribution 
$P(l)$ of the widths $l$ of effective barrier regions. [We normalise
$P(l)$ to $\int_0^\infty \mathrm{d}l\, lP(l)=1$, such that the
probability for a randomly chosen transition state to be in a barrier
region of width $l$ is $lP(l)$. Results for $P(l)$ from our current RG
scheme are not given here because even for a pure trap
model the RG flow eventually leads to a mixture of effective traps and
effective barriers which cannot be characterised by a single
distribution $P(l)$.]
If a given barrier region is delimited by sites $b$ and 
$b+l$, then the propagator is $G_{n,m} = \delta_{n,b} \frac{b+l-m}{l}
+ \delta_{n,b+l} \frac{m-b}{l}$ for $b\leq m\leq b+l$.  Averaging
over all initial sites $m$ at fixed $\rr=n-m$ on a long chain with the relevant
distribution of barrier widths, we arrive at
the disorder-averaged diffusion front,
\begin{equation}
\overline{G_\rr} = \int_{|\rr|}^\infty \mathrm{d}l\, \frac{l-|\rr|}{l} P(l)
\label{equ:gl}
\end{equation}
where we have assumed that $t$ is large, so that 
$P(l)$ is smooth, and we may convert sums over $l$ to integrals.
Within the Mon03 scheme, the distances $l$ are all independently
and identically distributed with an exponential form
$P(l)=\xi^{-2} \ee^{-l/\xi}$.  This leads to an estimate
for the diffusion front in the limit of large $\mu$:
\begin{equation}
\overline{G_\rr} \approx \xi^{-1} \ee^{-\rr/\xi} \int_0^\infty\mathrm{d}y 
  \frac{y}{y+\rr/\xi} \ee^{-y}
\label{equ:mon_diff}
\end{equation}
so that the diffusion front is a scaling function of $x=\rr/\sigma$ 
for large times, with $\sigma=\xi$.
The assumption of Ref.~\cite{Mon03} is that, while working at large
finite $\mu$ does affect $P(l)$, 
these changes lead to perturbative corrections to the diffusion front.

However, Fig.~\ref{fig:barr_eff} 
shows that the tail of the diffusion front is rather different
from the prediction (\ref{equ:mon_diff}) of Ref.~\cite{Mon03}, at least for 
$\mu=3$.  The effective dynamics scheme discussed in this paper gives more  
accurate predictions for this tail.  In fact, the convergence
of the tail of the diffusion front to its large-$\mu$ prediction
is quite slow.  Instead of plotting the diffusion front data directly,
we show barrier width distributions 
$P(l)$ for the effective dynamics schemes of Refs.~\cite{Mon03}
and~\cite{JS-dual}. (The distribution $P(l)$ is obtained
directly from the effective dynamics: in fact this was the route
by which $\Gbar_\rr$ was evaluated in Figs.~\ref{fig:fronts}
and~\ref{fig:barr_eff}(a).)
If we define the mean barrier width
to be $\xi$, then we find that $P(l)$ converges to the simple
exponential distribution only if we take $\mu\to\infty$ at
a fixed value of the scaling variable $l/\xi$.  For 
smaller $l$ there are corrections to this
distribution that cannot be accounted for by treating $1/\mu$
perturbatively.
Indeed, our numerical results are most consistent
with the tail of the diffusion front scaling as
\begin{equation}
\log \overline{G_\rr} \sim -|\rr|^{1+\alpha} 
\end{equation} 
where $\alpha>0$ is a power that vanishes as $\mu\to\infty$.  In this
case it is clear that the limits of large $\mu$ and large $\rr$ do
not commute and hence that working perturbatively in $1/\mu$
is likely to fail when considering the large-$\rr$ limit of $\overline{G_\rr}$.
We have
not found an analytical treatment which can determine
the resulting exponent $\alpha$, neither exactly nor for our effective
dynamics. 
However, the numerical evidence of Fig.~\ref{fig:barr_eff}
is that our scheme does capture the
non-perturbative effects which lead to slow convergence of the
diffusion front to the large $\mu$ limit.

\section{Outlook}
\label{sec:outlook}

In this article, we have derived
duality relations that connect pairs of hopping models linked by an
inversion of their energy landscape.
The simplest case is that of models with equilibrium steady states and
periodic boundary conditions, but we were able also to link models with
absorbing and reflecting boundary conditions.
Somewhat surprisingly, certain
periodic systems (pure trap and barrier models with a bias) in which
the steady state has a finite current can be analysed
similarly~\cite{forthcoming}. 
All duality relations are initially expressed in terms of the relevant
master operators, but we showed that one can then also construct the
propagator of each model from its dual. It follows further that the 
disorder-averaged propagators in
each pair of models are equal on all time and length scales.
We discussed an alternative duality relation giving the same results,
which is independent of the disorder and related to one used 
by Sch\"utz and Mussawisade~\cite{schutz} for a reaction-diffusion
model.

We have also introduced an effective dynamics scheme for
these hopping models.  It incorporates both the scheme of Ref.~\cite{JS-dual}
and that of DMF, allowing a broad class of models to be treated in a unified
fashion.  For a range of ``mixed trap-barrier models'', including 
the pure barrier and trap
cases, we have also shown that our scheme captures non-perturbative
corrections to the schemes of DMF and Mon03.

Our results also identify a few questions: 
can explicit expressions for 
the disorder-averaged diffusion fronts be derived for the
mixed model or for the pure trap/barrier cases, either exactly
or at least within the effective dynamics?  It appears that the
diffusion front in the mixed model depends only on the larger
of $\mub$ and $\mut$: can this be established?  More speculatively,
one might ask if the
methods used here can be generalised in order to identify effective
trap and barrier regions for higher-dimensional systems. We leave
these issues for future work.

\ack
We thank Jean-Philippe Bouchaud, Jeppe Dyre, Jorge Kurchan,
Peter Mayer and C\'ecile Monthus 
for helpful discussions.
We thank Gunter Sch\"utz for bringing Ref.~\cite{schutz} to our attention.

\begin{appendix}

\section{Effective dynamics}
\label{app:eff}

In this appendix we give some details of the renormalisation
scheme that underlies our effective dynamics.  The scheme gives a good
description of the dynamics of the model in the limit in which
rates in the same neighbourhood are sufficiently well-separated.
We first derive the version of the RG procedure that is most natural
from a formal point 
of view, but then argue in favour of the more physically-motivated
scheme described in the main text.  Both schemes agree
in the relevant limit where time scales are locally
well-separated, and we argue that the scheme of the main text
captures the subleading corrections to this limit more effectively.

\subsection{Formal scheme}
\label{app:basic}

Suppose that we have a master operator $W_\mathrm{R}(\Gamma)$ of the
form given in (\ref{equ:wr_def}), and that this operator
gives an accurate description of motion on time scales
longer than $\Gamma^{-1}$.  We wish to construct a projection
operator $P(\Gamma-\delta\Gamma)$ which represents
a good approximation to the operator $P_\mathrm{ex}(\Gamma-\delta\Gamma)$
of (\ref{equ:Pex}), so that $W_\mathrm{R}(\Gamma-\delta\Gamma)
=P(\Gamma-\delta\Gamma)W_\mathrm{R}(\Gamma)P(\Gamma-\delta\Gamma)$
gives an accurate description of motion on time scales
longer than $(\Gamma-\delta\Gamma)^{-1}$.  In addition,
for the scheme to represent a renormalisation group flow,
we require that $W_\mathrm{R}(\Gamma-\delta\Gamma)$ is
also of the form given in (\ref{equ:wr_def}).

As discussed in the text,
we begin by estimating an eigenmode of $W_\mathrm{R}(\Gamma)$ that
is concerned with fast motion.
To achieve this, we imagine that
all the rates $r_\alpha$ and $\ell_\alpha$ are associated
with very slow motion, except the triplet $\{ r_\alpha, \ell_\alpha,
\ell_{\alpha+1} \}$.  In that case we can write
the master operator as
\begin{equation}
\fl W_\mathrm{R}(\Gamma)
 \approx W_0 = \left(|P_{\alpha-1}\rangle\quad |P_\alpha\rangle \quad
                    |P_{\alpha+1}\rangle \right) 
\left( \begin{array}{ccc}
0 & \ell_\alpha & 0 \\
0 & -(\ell_\alpha+r_\alpha) & \ell_{\alpha+1} \\
0 & r_\alpha & -\ell_{\alpha+1} \end{array} \right) 
\left( \begin{array}{c} \langle Q_{\alpha-1} | \\
                        \langle Q_\alpha | \\ \langle Q_{\alpha+1}|
\end{array} \right) 
\label{equ:fast}
\end{equation}
Diagonalising yields three eigenvalues.  Since we have assumed
that transitions out of site $\alpha-1$ are very slow
the `steady state', i.e.\ the right eigenvector with eigenvalue zero,
of this reduced system is simply localised 
on that site.  Then there are two negative eigenvalues whose
moduli are
$\rho_\pm
=
 \hf(r_\alpha + \ell_\alpha + \ell_{\alpha+1})
\pm \hf \sqrt{ r_\alpha^2 + 2 r_\alpha ( \ell_\alpha + \ell_{\alpha+1} )
 + (\ell_\alpha-\ell_{\alpha+1})^2 }$.  These eigenvalues are
associated with fast ($+$) and slow ($-$) motion.  We identify
$\rho_+$ as a rate for fast motion to the right
from effective trap $\alpha$.
The effective dynamics proceeds by successive removal
of the fastest such modes: recall (\ref{equ:lam2}),
where $\rho_+$ is written as $\rho_\alpha$.
As discussed in the main text in addition to triplets of rates
$(r_\alpha, \ell_\alpha, \ell_{\alpha+1})$, we also consider triplets
such as $(\ell_\alpha, r_\alpha, r_{\alpha-1})$, for which the same
treatment applies, with the rate $\lambda_\alpha$ of the fastest mode given in
(\ref{equ:lam1}). In the discussion below we assume for 
concreteness that the largest
approximate eigenvalue among the $\{\rho_\alpha,\lambda_\alpha\}$ is
$\rho_\alpha\equiv \rho_+$.

To accomplish the removal of the fastest mode, we will
project the original master operator
$W_\mathrm{R}(\Gamma)$ onto the basis spanned by the eigenvectors
associated with the slow motion.  
More precisely, 
the zero eigenvectors of the approximate master operator $W_0$
are $|P_{\alpha-1}\rangle$ to the right
and $\langle e_3| = \langle Q_{\alpha-1} | +
\langle Q_\alpha | + \langle Q_{\alpha+1}|$ to the left. The 
right and left slow eigenvectors, corresponding to eigenvalue
$\rho_-$, we write as $|\rho_-\rangle$
and $\langle \rho_-|$.
We therefore define a projection operator whose matrix
elements will give the propagator on time scales
longer than $1/\rho_+$:
\begin{equation}
\fl {\cal P}_+ \equiv 
\sum_{\alpha'=-\infty}^{\alpha-2}
|P_{\alpha'}\rangle \langle Q_{\alpha'}| 
+
|P_{\alpha-1}\rangle \langle e_3|
+
|\rho_- \rangle \langle \rho_-|
+
\sum_{\alpha'=\alpha+2}^{\infty}
|P_{\alpha'}\rangle \langle Q_{\alpha'}| 
\label{equ:proj_exact}
\end{equation}

In general, the operator $\mathcal{P}_+ W_\mathrm{R}(\Gamma) \mathcal{P}_+$
is not of the same form as $W_\mathrm{R}(\Gamma)$:
it contains hopping processes between next-nearest neighbours for
the effective traps, and does not represent a suitable
approximation to $W_\mathrm{R}(\Gamma-\delta\Gamma)$.  
This effect is familiar in renormalisation schemes, and requires
irrelevant terms in the master equation to be discarded. In our situation,
next-nearest neighbour hopping becomes irrelevant as time scales
become well-separated.  For example, in the barrier-like case where 
$r_\alpha\approx \ell_{\alpha+1}$ but
$r_\alpha\gg \ell_\alpha$, we have
\begin{eqnarray}
\fl {\cal P}_+ \approx {\cal P}_\mathrm{B} &\equiv&
\sum_{\alpha'=-\infty}^{\alpha-2}
|P_{\alpha'}\rangle \langle Q_{\alpha'}| 
+
|P_{\alpha-1}\rangle \langle Q_{\alpha-1}|
+
|P'\rangle (\langle Q_{\alpha}| + \langle Q_{\alpha+1}|)
\nonumber \\ \fl & &
+
\sum_{\alpha'=\alpha+2}^{\infty}
|P_{\alpha'}\rangle \langle Q_{\alpha'}|  
\label{equ:Pbarr}
\end{eqnarray}
with 
\begin{equation}
|P'\rangle=\frac{\ell_{\alpha+1}}{r_\alpha+\ell_{\alpha+1}}|P_\alpha\rangle
+\frac{r_{\alpha}}{r_\alpha+\ell_{\alpha+1}}|P_{\alpha+1}\rangle.
\label{equ:Pnew}
\end{equation}  
We identify $|P'\rangle$ as the right 
eigenvector associated with a new effective trap
that combines traps $\alpha$ and $\alpha+1$.  The resulting
master operator $\mathcal{P}_\mathrm{B} W_\mathrm{R}(\Gamma)
 \mathcal{P}_\mathrm{B}$ 
is now of the same form as $W_\mathrm{R}(\Gamma)$, 
with the free energy of the new effective trap 
given by $\ee^{F_{\alpha,\alpha+1}} = \ee^{F_\alpha}+\ee^{F_{\alpha+1}}$.
Thus, we may perform an RG step by taking
$\mathcal{P}(\Gamma-\delta\Gamma)=P_\mathrm{B}$ as our approximation
to $\mathcal{P}_\mathrm{ex}(\Gamma-\delta\Gamma)$.
Such an RG step is illustrated by the transition between
Fig.~\ref{fig:rg_new}b and Fig.~\ref{fig:rg_new}c: consistent
with that figure, the transition
states on either side of the new trap are unchanged during
this procedure.

On the other hand, in the trap-like case where $\ell_\alpha\approx
 r_\alpha$ but $r_\alpha\gg\ell_{\alpha+1}$, we have
\begin{eqnarray}
\fl {\cal P}_+ \approx {\cal P}_\mathrm{T} &\equiv&
\sum_{\alpha'=-\infty}^{\alpha-2}
|P_{\alpha'}\rangle \langle Q_{\alpha'}| 
+
|P_{\alpha-1}\rangle (\langle e_3| - \langle Q'|)
+
|P_{\alpha+1}\rangle \langle Q'|
\nonumber \\  \fl & & 
+
\sum_{\alpha'=\alpha+2}^{\infty}
|P_{\alpha'}\rangle \langle Q_{\alpha'}|  
\label{equ:Ptrap}
\end{eqnarray}
with
\begin{equation}
\langle Q'|=
\frac{r_\alpha}{r_\alpha+\ell_\alpha}\langle Q_\alpha|+\langle Q_{\alpha+1}|.
\label{equ:Qnew}
\end{equation}  
As for the previous case, $\mathcal{P}_\mathrm{T} W_\mathrm{R}(\Gamma)
\mathcal{P}_\mathrm{T}$ is of the same form as $W_\mathrm{R}(\Gamma)$,
so the choice ${\cal P}(\Gamma-\delta\Gamma)={\cal P}_\mathrm{T}$
corresponds to a valid RG step.
In this step, trap $\alpha$ has been incorporated
into a new effective barrier region that merges the old barrier
regions $\alpha-\hf$ and $\alpha+\hf$. Its escape
properties to the remaining effective traps $\alpha+1$ and $\alpha-1$
are described by the eigenvectors $\langle Q'|$ and $\langle e_3| -
\langle Q'|=\langle
Q_{\alpha-1}|+\ell_\alpha/(r_\alpha+\ell_\alpha)\langle Q_\alpha|$,
respectively. 
An example of such an RG step takes place between
Figs.~\ref{fig:rg_new}a and~\ref{fig:rg_new}b. 

It should be noted that if $r_\alpha$ is much greater than both $\ell_\alpha$
and $\ell_{\alpha+1}$ then both (\ref{equ:Pbarr}) and (\ref{equ:Ptrap}) 
reduce to the case of DMF, which is
$|P'\rangle=|P_{\alpha+1}\rangle$
and $\langle Q'|=\langle Q_{\alpha}| + \langle Q_{\alpha+1}|$.
In practice, for any renormalisation step, we choose either to
combine traps by taking $P(\Gamma-\delta\Gamma)=P_\mathrm{B}$
or otherwise to combine barriers by taking 
$P(\Gamma-\delta\Gamma)=P_\mathrm{T}$.
In either case the resulting
$W_\mathrm{R}(\Gamma-\delta\Gamma)$ is
indeed of
the same form as $W_\mathrm{R}(\Gamma)$: this ensures
that our procedure is a valid  renormalisation flow in the space of 
hopping models.  

As discussed in Sec.~\ref{sec:separation},
the validity of the renormalisation 
scheme requires that the projection operator
evolves with $\Gamma$ such that
$\mathcal{P}(\Gamma-\delta\Gamma)\approx 
 \mathcal{P}_\mathrm{ex}(\Gamma-\delta\Gamma)$.
Assuming that we combine traps $\alpha$ and $\alpha+1$, we should
have ${\cal P}_B \approx {\cal P}_+ \approx {\cal P}_\mathrm{ex}$.
Applying perturbation theory to the fast eigenvectors, we find
that corrections are small if 
$r_\alpha+\ell_{\alpha+1}\gg r_{\alpha-1}, \ell_\alpha,
r_{\alpha+1}, \ell_{\alpha+2}$.  
Similarly, if we combine
barriers $\alpha\pm\hf$, we require
$\ell_\alpha+r_\alpha\gg \ell_{\alpha-1}, r_{\alpha-1}, \ell_{\alpha+1},
r_{\alpha+1}$.  Essentially, if the rate $r_\alpha$ is large at
a given stage of the RG
(in the sense that $\rho_\alpha>\Gamma-\delta\Gamma$), then 
$r_\alpha$ must be larger than all rates in the neighbourhood, 
except for either $\ell_\alpha$ or $\ell_{\alpha+1}$, one
of which may be comparable to $r_\alpha$.

\subsection{Illustrative example, and comparison with physical scheme of
the main text}

The formal scheme derived above differs from the one set out in the main text
in the way effective traps and barriers are combined. 
We illustrate this with the system of Fig.~\ref{fig:rg_new}.
The bare master operator is
\begin{equation}
\fl
W=\cdots 
+ (|0\rangle - |1\rangle)(\ell_1 \langle 1| - r_0\langle 0|) 
+ (|1\rangle - |2\rangle)(\ell_2 \langle 2| - r_1\langle 1|) 
  + \cdots 
\end{equation}
where the $\cdots$ indicate the remaining terms in the master
operator, including those for hopping into and out of
this segment of the chain.

Applying the first step of the formal RG scheme, 
we construct the projection operator 
\begin{equation}
\mathcal{P} = \cdots + |0\rangle \langle Q_0| + |2\rangle \langle Q_2|
+\cdots
\label{equ:peg1_new}
\end{equation}
where $\langle Q_0|=\langle 0| + \ee^{E_\hf-F_{\barr}}\langle 1|$ 
and $\langle Q_2|=\langle 2| + \ee^{E_{\frac32}-F_{\barr}}\langle 1|$
are constructed according to (\ref{equ:Qnew}),
with $\ee^{F}=\ee^{E_{\frac12}}+\ee^{E_\frac32}$.
This leads to the same result as the physical scheme
of the text, since (\ref{equ:peg1_new}) is also
consistent with (\ref{equ:Qrg}) above.
The renormalised master operator is
\begin{equation}
W_\mathrm{R} = \mathcal{P} W \mathcal{P} =
\cdots + (|0\rangle - |2\rangle) \ee^{-F_{\barr}}
(\ee^{-E_2} \langle Q_2| - \ee^{-E_0} \langle Q_0|)
+\cdots
\label{equ:weg1_new}
\end{equation}

Applying the formal scheme again to this operator, the projection
operator is 
\begin{eqnarray}
\mathcal{P}' 
&=& \cdots + |P_{02}\rangle ( \langle Q_0| + \langle Q_2| ) + \cdots
\nonumber\\
&=& \cdots + |P_{02}\rangle ( \langle0| + \langle1| + \langle2| ) + \cdots
\label{equ:peg2_new}
\end{eqnarray}
where $|P_{02}\rangle=\ee^{-F_{02}} ( \ee^{E_0} |0\rangle + \ee^{E_2} |2\rangle)$
with $\ee^{F_{02}}=\ee^{E_0}+\ee^{E_2}$.
According to the scheme of the main text,
we would have obtained a similar result, but with the replacement
$|P_{02}\rangle\to|P_{012}\rangle=\ee^{-F_{012}} ( \ee^{E_0} |0\rangle + 
\ee^{E_1} |1\rangle + \ee^{E_2} |2\rangle)$,
where $\ee^{F_{012}}=\ee^{E_0}+\ee^{E_1}+\ee^{E_2}$.

In the limit where rates are well-separated and the formal scheme
is is exact, it follows from~(\ref{equ:eg_rates}) that
$E_1\ll E_0,E_2$, and in this case, the physical and formal
schemes coincide.  Indeed, it may be shown
that the errors associated with
the physical and formal schemes are of the same
order.
We conclude that the physical scheme
is at least as appropriate as the formal one and indeed
it can be verified that the errors are of the same order.

The key point is that the physical scheme of the text
is based on the assumption of equilibration within effective
traps.  Under that assumption, it is clear that the
states $|P_\alpha\rangle$ should coincide with
Boltzmann distributions over the sites within the trap.
In general, the formal scheme gives a state $|P_\alpha\rangle$
that is finite only on a subset of the sites within the trap.
This is a necessary feature of the formal scheme, 
because the $|P_\alpha\rangle$ at each stage are constructed
from the $|P_\alpha\rangle$ of the previous stage.  In the language
of section~\ref{sec:rg}, the slow co-ordinates $p_\alpha(t)$
must be linear combinations of the $p_\alpha$ of the previous
stage.  Thus, since site $1$ in the example of Fig.~\ref{fig:rg_new}
is not contained in any of the $p_\alpha$ after the first
step of the RG, it can never be part of an effective trap at any future
stage.  Equivalently, the $|P_\alpha\rangle$ can have no
contribution from site 1 in future stages, and may not correspond
to Boltzmann distributions of the whole trap.  This in turn
means that the approximate propagator 
$\langle n|\mathcal{P}(\Gamma)|m\rangle$ is equal to
zero for some final sites $n$ within effective traps,
underestimating the true value of $G_{nm}(\Gamma^{-1})$ there.
The differences between the formal and physical schemes
are shown numerically in Fig.~\ref{fig:mix-num}, which illustrates how
the formal scheme underestimates the probability of
propagation onto certain sites within the effective traps, for specific
disorder realisations.  As discussed in Sec.~\ref{sec:numerics},
the physical and formal schemes show differences in the tails
of the disorder-averaged diffusion front that are of the same
order as the deviations between real and effective dynamics.

Finally, we note that the physical scheme is not strictly a 
renormalisation flow in that the rates and free energies at a given 
stage, depend not just on the renormalised operator $W_\mathrm{R}$
at the previous stage, but on all the bare energies
$E_i$ and $E_{i+\hf}$.  If this feature is considered
undesirable, the formal scheme may be used.  However, the interpretation
of the physical scheme as a partition into contiguous
trap and barrier regions and the intuitive idea of equilibration
within effective trap regions means that we prefer that route.

\end{appendix}

\end{document}